\documentclass[11pt]{amsart}

\usepackage{amsmath, amsthm, amssymb} 
\numberwithin{equation}{section}

\usepackage{amsthm}
\usepackage{hyperref}
\usepackage{youngtab}
\usepackage{mathabx}
\usepackage[all]{xy}
\usepackage[usenames,dvipsnames]{xcolor}

\theoremstyle{indented}

\usepackage[left=2.5cm,right=2.5cm,top=3.5cm]{geometry}

\pdfoutput=1
\usepackage{amsmath}
\usepackage{amssymb}
\usepackage{wrapfig}
\usepackage{bbm}
\usepackage{arydshln}
\usepackage{multirow}
\usepackage{mathtools}
\usepackage{blkarray}
\usepackage{setspace}
\usepackage{dsfont}
\usepackage{float}
\usepackage{subcaption}
\usepackage{braket}
\usepackage[utf8]{inputenc}
\usepackage[english]{babel}
\usepackage[all]{xy}
\usepackage{tikz-cd}
\usepackage{mathtools}
\usepackage{tikz}
\usepackage{comment}
\usepackage{amssymb}
\usepackage{graphicx}
\usepackage{subcaption}

\makeatletter
\newcommand\mathcircled[1]{%
  \mathpalette\@mathcircled{#1}%
}
\newcommand\@mathcircled[2]{%
  \tikz[baseline=(math.base)] \node[draw,circle,inner sep=3pt] (math) {$\m@th#1#2$};%
}
\makeatother

\allowdisplaybreaks

\begin{document}

\title{Branes, Quivers and BPS Algebras}

\author{Miroslav Rap\v{C}\'{a}k}

\begin{abstract}
These lecture notes cover a brief introduction into some of the algebro-geometric techniques used in the construction of BPS algebras. The first section introduces the derived category of coherent sheaves as a useful model of branes in toric Calabi-Yau three-folds. This model allows a rather simple derivation of quiver quantum mechanics describing low-energy dynamics of various brane systems. Vacua of such quantum mechanics can be identified with the critical equivariant cohomology of the moduli space of quiver representations. These are often counted by various crystal configurations. Using correspondences in algebraic geometry, one can construct rich families of affine-Yangian representations. We conclude with an exploration of different algebraic structures naturally appearing in our story. The material was covered in a 4-lecture mini-course within the Second PIMS Summer School on Algebraic Geometry in High-Energy Physics. The text contains some new ideas, examples and remarks that are going to be covered in detail in a joint work with Dylan Butson.
\end{abstract}

\maketitle

\tableofcontents

\section{Physical motivation}

These lecture notes cover a brief introduction into some of the algebro-geometric techniques used in the construction of BPS algebras. Given the limited space together with the presence of a few minor loopholes that require further investigation, it would be impossible to provide a rigorous exposition of the full story. On the other hand, if we restricted only to a subset of the presented material, we would not be able to appreciate the beauty of the whole construction. I have  thus decided to strip off as many formalities as possible and simply illustrate the main ideas, concepts and techniques in a few examples. I hope the discussion below to serve as a motivation for diving deeper into the subject of BPS algebras and geometric representation theory.

Our starting point is the ten-dimensional type-IIA string theory together with its D0-, D2-, D4-, D6- and D8-branes of dimensions 1, 3, 5, 7 and 9 respectively. String theory is heavily used to geometrically engineer supersymmetric quantum field theories via the process of compactification \cite{Aspinwall:1995xy,Kachru:1995wm,Katz:1996fh,Klemm:1996bj,Katz:1997eq}. Imagine we are interested in some theory living on a space that is a product of two manifolds. If one of the factors is very small, the full theory should admit an effective description in terms of a lower-dimensional theory living on the remaining large factor of the space.  For example, studying string theory on $M_4 \times M_6$, with the subscript labeling the dimension of the manifold, and sending the volume of $M_6$ to zero, we expect the system to have an effective description in terms of a theory on $M_4$. The resulting theory obviously depends on the geometry of the compactification manifold $M_6$. In order for the resulting theory to preserve some supersymmetry, the compactification manifold needs to satisfy further constraints. For example, to engineer a four-dimensional theory on $M_4$ with an $\mathcal{N}=2$ supersymmetry, we require $M_6$ to be a Calabi-Yau threefold. 

Geometric properties of $M_6$ capture a lot of information about the resulting four-dimensional theories. For example, supersymmetric (BPS) operators can be engineered from D-branes wrapping various complex submanifolds inside $M_6$. In particular, BPS particles and line operators (one dimensional objects in $M_4$) arise from D0-branes supported at a point in $M_6$, D2-branes wrapping a two-cycle in $M_6$, D4-branes wrapping a four-cycle in $M_6$ or D6-branes wrapping the full compactification manifold $M_6$. 

In our discussion, we are going to restrict to the simplest six-dimensional Calabi-Yau manifold $M_6=\mathbb{C}^3$. Most of our discussion has (or is expected to have) a generalization to more complicated Calabi-Yau threefolds but their discussion goes beyond the scope of this note. We are going to see that already this trivially-looking example leads to an enormously rich story.

One might raise an objection that $\mathbb{C}^3$ is not a compact manifold and the whole construction suggested above is meaningless. Luckily, one can introduce a deformation of the theory (by turning on a particular background vacuum expectation value for the B-field \cite{Hellerman:2011mv,Hellerman:2012zf,Costello:2016nkh}) known as the $\Omega$-background \cite{Moore:1997dj,Moore:1998et,Dorey:2002ik,Nekrasov:2002qd} and parametrized by\footnote{We are actually going to impose condition $\epsilon_1+\epsilon_2+\epsilon_3=0$ specializing to the subtorus preserving the Calabi-Yau volume form $dz_1\wedge dz_2\wedge dz_3$.} $\epsilon_1,\epsilon_2,\epsilon_3$ associated with the three $U(1)$ actions rotating the three coordinate lines $\mathbb{C}$ inside $\mathbb{C}^3$. Such an $\Omega$-deformation localizes the theory to the fixed-point of the $U(1)^3$ action (the origin) and effectively compactifies the theory to four dimensions. Whenever I want to stress the presence of the $\Omega$-deformation, I am going to write $\mathbb{C}^3$ as $\mathbb{C}_{\epsilon_1}\times\mathbb{C}_{\epsilon_2}\times\mathbb{C}_{\epsilon_3}$.

The introduction of the $\Omega$-background forces the support of the D-branes to be preserved by the (complexified) $U(1)^3$ action. Possible orientations of branes consistent with the $\Omega$-background are show in table \ref{branes}.
\begin{table}[h]
    \centering
    \begin{tabular}{|c|c | c c c |}\hline
        Type IIA & $M_4$ & $\mathbb{C}_{\epsilon_1}$ & $\mathbb{C}_{\epsilon_2}$& $\mathbb{C}_{\epsilon_3}$ \\ \hline
    D0   & $L$  &  & &   \\  \hline
    D2   & $L$  & $\times$ & &   \\
    D2   & $L$  &  & $\times$ &   \\
    D2   & $L$  &  & & $\times$  \\  \hline
    D4   & $L$  &  & $\times$ & $\times$   \\
    D4   & $L$  &$\times$  & & $\times$   \\
    D4   & $L$  &  $\times$  &$\times$ &  \\  \hline
    D6   & $L$  & $\times$ & $\times$ & $\times$  \\ \hline
    \end{tabular}
\caption{Possible support of branes in $\Omega$-deformed $\mathbb{C}^3$ wrapping a line $L$ inside $M_4$.}
\label{branes}
\end{table}

Given a non-compact Calabi-Yau manifold, we can distinguish branes according to the compactness of their support. Branes with compact support are going to be treated as light, dynamical objects in the compactified theory (BPS particles). On the other hand, branes with non-compact support are going to be heavy, non-dynamical line operators.  In our example of $M_6=\mathbb{C}^3$ in the presence of the $\Omega$-background, the only compact brane is a D0-brane.\footnote{This is generally not going to be the case if the manifold cointains compact two-cycles or even four-cycles. For example, the topal space of $\mathcal{O}(-1)\oplus \mathcal{O}(-1)\rightarrow \mathbb{CP}^1$ is a toric Calabi-Yau manifold admitting a two-parametric $\Omega$-background and a D2-brane wrapping the zero sectionof the bundle is obviously compact.}

More generally, one might be interested in configurations containing multiple branes. We can for example consider a stack of $N$ branes wrapping one of the cycles from the table \ref{branes}. We can also study configurations of intersecting stacks of branes. A configuration of branes is thus specified by assigning a multiplicity to each of the elementary cycles from table  \ref{branes} specifying the number of branes in the corresponding stack. 

A given system of branes (specified by the above multiplicities) can be further deformed in two different ways. One can either turn on a non-trivial vacuum expectation value for the Higgs field living on their support or one can form a non-trivial bound state of branes (possibly of different dimensions). We are going to see explicit examples of how to describe various brane configurations and their deformations in the next section. We are going to see that the natural language for describing branes in Calabi-Yau three-folds is provided by derived categories of coherent sheaves \cite{Kontsevich:1994dn,Sharpe:1999qz,Douglas:2000gi,Douglas:2000ah,Aspinwall:2001dz,Aspinwall:2001pu,Sharpe:2003dr,Aspinwall:2004bs,Aspinwall:2009isa}.

\begin{figure}
  \centering
    \includegraphics[width=0.47\textwidth]{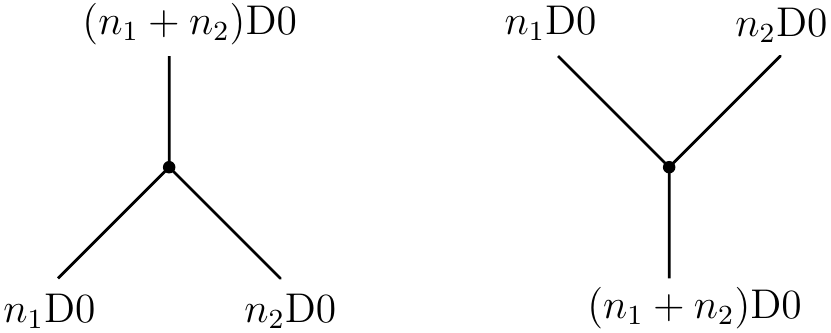}
\caption{Scattering of BPS particles arising from stacks of $n_1$, $n_2$ and $n_1+n_2$ D0-branes motivating the existence of BPS algebras.}
\label{fig1}
\end{figure}

As stated above, compactly supported branes give rise to particles of the compactified theory. Such particles can mutually scatter as depicted in figure \ref{fig1}. In particular, a particle associated with $n_1$ D0-branes and a particle associated with $n_2$ D0-branes can fuse to create a particle associated with a bound state of $n_1+n_2$ D0-branes. Analogously a bound state of  $n_1+n_2$ D0-branes can split into two bound states of $n_1$ and $n_2$ D0-branes respectively. Fixing a triple of vacua associated to the endpoints of the trivalent vertices from figure \ref{fig1}, one can compute its scattering amplitude. This gives rise to a multiplication structure on the space of such vacuum states
\begin{eqnarray}
|\lambda_1\rangle \otimes |\lambda_2\rangle \rightarrow |\lambda_3\rangle ,
\end{eqnarray}
where $|\lambda_1\rangle$ denotes a vacuum of the quantum mechanics arising from $n_1$ D0-branes,  $|\lambda_2\rangle$ is a vacuum of the quantum mechanics arising from $n_2$ D0-branes and  $|\lambda_3\rangle$ is a vacuum associated with  $n_1+n_2$ D0-branes. This picture motivates the existence of so-called BPS algebras as proposed in \cite{Harvey:1996gc}. In order to give a precise definition of BPS algebras, \cite{Kontsevich:2010px} introduced the concept of cohomological Hall algebra.\footnote{A precise connection of the cohomological Hall algebra with the physical picture of scattering amplitudes is far from being fully understood. For some attempts to relate these two structures, see \cite{2019aa}.} In the past couple of years, BPS algebras and their construction in terms of the cohomological Hall algebra received an enormous interest in both physics and mathematics. See for example \cite{Grojnowski:1995uda,Lusztig1991QuiversPS,Nakajima:1994nid,Nakajima:1995ka,Losev:1995cr,2011dd,schiffmann2012cherednik,Davison:2013nza,Rimanyi:2013yma,2013cc,Soibelman:2014gta,Rapcak:2018nsl,Dedushenko:2017tdw,Feigin:2018bkf,Ren:2015zua,2017ee,2019,Kapranov:2019wri,2019aa,2019bb,Rapcak:2020ueh,Li:2020rij,Galakhov:2020vyb,Toda:2020ysz,diaconescu2020mckay,Galakhov:2021xum,Galakhov:2021vbo,latyntsev2021cohomological,porta2021twodimensional}.

\begin{figure}
  \centering
    \includegraphics[width=0.5\textwidth]{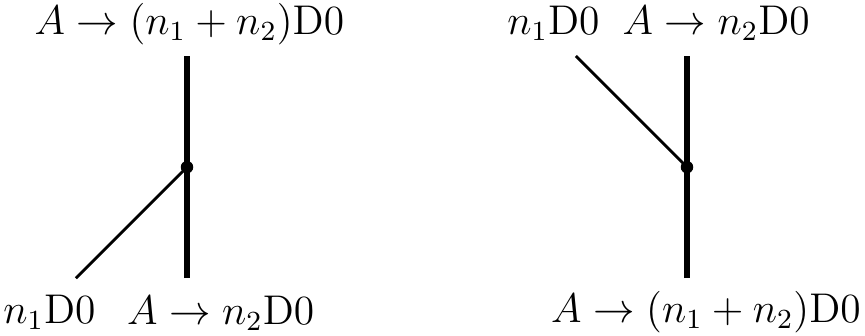}
\caption{Scattering of a BPS particle arising from a stack of $n_1$ D0-branes and a BPS line operator arising from a bound state of a non-compactly supported branes $A$ with $n_2$ D0-branes motivating the existence of modules for BPS algebras.}
\label{fig2}
\end{figure}

Let us fix a configuration of non-compact branes (say a stack of $N$ D4-branes along $\mathbb{C}_{\epsilon_1}\times \mathbb{C}_{\epsilon_2}$ inside $\mathbb{C}_{\epsilon_1}\times \mathbb{C}_{\epsilon_2}\times \mathbb{C}_{\epsilon_3}$) with $n$ D0-branes bound\footnote{This configuration is known to be a string-theory realization of instantons \cite{Witten:1994tz,Douglas:1995bn,Douglas:1996sw}.} to it. We are going to label such a configuration as $A\rightarrow n\ \mbox{D0}$, where $A$ specifies the configuration of non-compact branes. Analogously to the above, one can expect that processes of bounding/removing D0-branes as depicted in figure \ref{fig2} are going to lead to a module structure on the space of vacua associated with our bound states. If we define a weight of a configuration as the number of D0-branes then bounding D0-branes (increasing $n$) should lead to an action of raising operators whereas removal of D0-branes (decreasing $n$) should produce an action of lowering generators. This reasoning leads to a conjecture that a double of the cohomological Hall algebra (one copy increasing and one copy decreasing the number of D0-branes) should admit a module for any configuration of non-compactly-supported branes. In the simplest example of $\mathbb{C}^3$, the relevant double is known as the (shifted) affine Yangian of $\mathfrak{gl_1}$. This leads us to a correspondence \cite{Rapcak:2020ueh,BR}:
\begin{eqnarray}\nonumber
 \boxed{\mbox{Branes in}\ \mathbb{C}_{\epsilon_1}\times \mathbb{C}_{\epsilon_2}\times \mathbb{C}_{\epsilon_3}}  \quad \leftrightarrow \quad \boxed{\mbox{$\mathfrak{gl}_1$ affine Yangian modules}}
\end{eqnarray}
The aim of this note is to illustrate the construction of such modules on a few very simple examples. 

Our construction is going to proceed in the following three steps:
\begin{eqnarray}\nonumber
 \boxed{\mbox{Brane configuration}} \rightarrow \boxed{\mbox{Quiver quantum mechanics}}\rightarrow \boxed{\mbox{BPS states}} \rightarrow  \boxed{\mbox{Yangian module}}
\end{eqnarray}
Let me conclude this section by giving a few details about each of the steps we need to take on our journey towards BPS algebras: 
\begin{itemize}
\item \textbf{Brane configuration} $\rightarrow$ \textbf{Quiver QM:} The low-energy dynamics of a system of branes is expected to be described by a quantum field theory living on its support. That means that after compactification, the low-energy dynamics of our compactly-supported branes should be captured  supersymmetric quantum mechanics (QM) living on the one-dimensional line in the non-compact part of the space-time $M_4$. The field content of the effective field theory describing the dynamics of branes is usually determined by an analysis of string spectra in a prescribed background of branes. Here, we are not going to follow the standard physical derivation but rather use the fact that the relevant spectra can be determined by a relatively simple calculation within the derived category of coherent sheaves. The resulting field content is going to be encoded in a quiver diagram. We are going to see that the potential of such a theory can be derived easily in this context as well. The quiver with potential uniquely determines the desired supersymmetric quantum mechanics. Section 2 of this notes can be viewed as a concise overview of the material covered in \cite{Sharpe:2003dr} enriched by the discussion of potential from \cite{Aspinwall:2004bs} and a general framing from \cite{BR}. 
\item \textbf{Quiver QM} $\rightarrow$ \textbf{BPS vacua:} Throughout our discussion, are going to restrict to a subspace of so-called BPS vacua of our supersymmetric quantum mechanics. This subsector is characterized by living in the cohomology of one of the supercharges. We are going to identify this space (after a slight deformation of the quantum mechanics due to the presence of the non-trivial $\Omega$-background) with the equivariant critical cohomology of the moduli space of quiver representations. Working equivariantly allows us to identify the cohomology with fixed points of the corresponding moduli space lying in the critical locus of the potential. Counting such fixed points is going to lead to a rich combinatorics of melted crystals standard standard in the literature \cite{Okounkov:2003sp,Szendroi:2007nu,Pandharipande:2007sq,Mozgovoy:2008fd,Ooguri:2009ijd,Nagao:2010kx,Yamazaki:2010fz,Nishinaka:2013mba}. 
\item \textbf{BPS states} $\rightarrow$ \textbf{Yangian module:} Let us label the moduli space of vacua associated with $n$ D0 branes bound to a fixed configuration of non-compact branes $A$ as $\mathcal{M}(n)$. As we are going to see in section 4, for each $A$, there exists a correspondence $\mathcal{M}(n+1,n)$ with a map $p$ to $\mathcal{M}(n)$ and a map $q$ to $\mathcal{M}(n+1)$. Starting with an element in the (critical equivariant) cohomology $H^*(\mathcal{M}(n))$, pulling it back by $p^*$ and pushing forward by $q_*$, we are going to construct an action of raising operators of the BPS algebra.  Analogously, pulling back by $q_*$ and pushing forward by $p_*$ gives rise to the action of lowering generators of the algebra. See \cite{Nakajima:1994nid,Kontsevich:2010px,schiffmann2012cherednik,Soibelman:2014gta,2017ee,Rapcak:2020ueh} for details. In the example of branes in $\mathbb{C}^3$, the relevant BPS algebra is known as the $\mathfrak{gl}_1$ affine Yangian. Different configurations of branes $A$ then lead to various modules of the affine Yangian. Depending on the dimension of the support of our non-compact branes, we obtain very differently looking modules giving rise to various interesting algebraic structure. In particular, the geometric action is expected to factor through a map to Cherednik algebras (for D2-branes) \cite{Gaiotto:2020dsq,BR,RSYZ}, it factors through a map to corner vertex operator algebras (for D4-branes) \cite{Rapcak:2018nsl,Chuang:2019qdz,BR} and it gives rise to the MacMahon representations (for D6-branes) \cite{Rapcak:2020ueh}. The construction reproduces an extremely rich representation theory of the $\mathfrak{gl}_1$ affine Yangian and establishes its relation to geometry.
\end{itemize}

\section{Quivers from branes}

The first step in our construction is a derivation of quiver quantum mechanics describing a stack of D0-branes bound to different systems of D2-, D4- and D6-branes. The most straightforward yet tedious derivation would rely on the analysis of zero modes in the spectrum of open strings ending on involved branes \cite{Douglas:1996sw,Douglas:1995bn,Nekrasov:2016gud}. Instead of following this path, we are going to use the language of derived categories of coherent sheaves as a model of our D-branes and derive the quivers by studying morphisms in such a category. See \cite{Sharpe:2003dr} for an excellent introduction or \cite{Kontsevich:1994dn,kont,Sharpe:1999qz,Douglas:2000gi,Douglas:2000ah,Aspinwall:2001dz,Aspinwall:2001pu,Aspinwall:2004bs,ginz,derksen2008quivers,keller2009deformed,Aspinwall:2009isa,yan2014a,Bridgeland_t-structureson} for some of the original work.

\subsection{Derived category of coherent sheaves as a brane category}

Let us start with the introduction of objects in the derived category of coherent sheaves and justification that they form a good model for brane bound states. 

We start with a sketch of a formal definition of a sheaf. A sheaf $\mathcal{S}$ on a space $X$ is an assignment a space of sections $\mathcal{S}(U)$ to each open set $U$ together with a colection of restriction maps $\rho_{U,V}:\mathcal{S}(U) \rightarrow \mathcal{S}(V)$ for any $V\subset U$. These satisfy some obvious compatibility conditions:
\begin{enumerate}
\item  The inclusion $W\subset  V\subset U$ is consistent with the composition of restriction maps.
\item  The restriction map of $U\subset U$ is the identity.
\item For any pair $\sigma \in \mathcal{S}(U),\tau \in  \mathcal{S}(V)$ that agree on $U\cap V$, there exists a section $\rho \in  \mathcal{S}(U\cup V)$ that restricts to $\sigma$ and $\tau$ respectively.
\item If $\sigma \in \mathcal{S}(U\cup V)$ such that $\sigma|_U=\sigma|_{V}=0$, then $\sigma=0$.
\end{enumerate}

An example of a sheaf that is going to play an important role in our discussion is the structure sheaf $\mathcal{O}_X$ of a complex variety $X$. The structure sheaf  assigns the ring of algebraic functions on $U$ to each open set $U$. More generally, for any holomorphic bundle $E\rightarrow X$ of rank $k$, we can define a sheaf assigning the space of its holomorphic sections over $U$ to each open set $U$. Such a sheaf is obviously a module for the structure sheaf $\mathcal{O}_X$ since we can multiply sections of a bundle by functions. The class of sheaves of this form go under the name locally-free sheaves since they locally look like $\mathcal{O}_X^k|_{U}$. Since a brane in string theory is specified by its support together with a (Chan-Paton) bundle over it, it is natural to identify locally-free sheaves of rank $k$ with a stack of $k$ D6-branes wrapping the whole $X$. In our simplest example of $\mathbb{C}^3$, the structure sheaf assigns the coordinate ring
\begin{eqnarray}
\mathbb{C} [x_1,x_2,x_3]
\end{eqnarray}
to the whole $\mathbb{C}^3$ and it generally  assigns the ring of algebraic functions on $U$ to any other open subset $U$. Due to the triviality of the example at hand, all the locally-free sheaves are of the form $\mathcal{O}_X^{k}$ and they are actually free (not only locally-free).

Coherent sheaves form a class of sheaves that can be defined locally by imposing a set of relations on a locally-free sheaf, i.e. they can be locally identified with the cokernel of some map
\begin{eqnarray}
f:\mathcal{O}^l_X|_U\rightarrow \mathcal{O}^m_X|_U
\end{eqnarray}
for some integers $l,m$.

Let us now analyze some examples and argue that we can naturally associate a quasi-coherent sheaf to any of the branes discussed in the introduction. Since our simple example of $\mathbb{C}^3$ does not have any non-trivial global structure, we are going to write all the maps simply in the open set being the whole $U=\mathbb{C}^3$. For a trivial map
\begin{eqnarray}
f:0\rightarrow \mathbb{C} [x_1,x_2,x_3]^{ k},
\end{eqnarray}
the sheaf is formed by $k$ copies of the structure sheaf and can be identified with the stack of $k$ D6-branes wrapping $\mathbb{C}^3$ as discussed above.

Another example is the skyscraper sheaf at the origin that can be expressed as a cokernel of the following map
\begin{eqnarray}
\mathbb{C} [x_1,x_2,x_3]^3\xrightarrow{(x_1,x_2,x_3)}\mathbb{C} [x_1,x_2,x_3]=\mathbb{C},
\end{eqnarray}
i.e.
\begin{eqnarray}
\frac{\mathbb{C} [x_1,x_2,x_3]}{(x_1,x_2,x_3)}.
\end{eqnarray}
Away from the origin, the cokernel is obviously trivial since any $f\in \mathbb{C} [x_1,x_2,x_3]$ can be locally written as
\begin{eqnarray}
f=x_1 \left (\frac{f}{x_1}\right )
\end{eqnarray}
if $x_1\neq 0$ and analogously if we are away from $x_2=0$ and $x_3=0$. On the other hand, this is not possible at $x_1=x_2=x_3=0$, where the cokernel is one-dimensional. The corresponding sheaf is thus supported at the origin and it is natural to associated it with the D0-brane. To get a sheaf associated with the stack of $n$ D0-branes, we can simply take $n$ copies of such a sheaf
\begin{eqnarray}
\left (\frac{\mathbb{C} [x_1,x_2,x_3]}{(x_1,x_2,x_3)}\right )^{n}.
\end{eqnarray}

Analogously, for the map 
\begin{eqnarray}
\mathbb{C} [x_1,x_2,x_3]\xrightarrow{(x_3)}\mathbb{C} [x_1,x_2,x_3]=\mathbb{C}[x_1,x_2],
\end{eqnarray}
the cokernel gives a module
\begin{eqnarray}
\mathbb{C}[x_1,x_2,x_3]/(x_3)
\end{eqnarray}
that can be associated with a D4-brane supported along $x_3=0$. A sheaf associated with a multiple of the same support D4-branes is then a direct sum of $k$ copies of this module. One can similarly define a sheaf associated to D4-branes of other orientations by exchanging $x_3\rightarrow x_1,x_2$ and associate a sheaf to an arbitrary configuration of intersecting D4-branes by taking a direct sum.

The map 
\begin{eqnarray}
\mathbb{C} [x_1,x_2,x_3]^2\xrightarrow{(x_1,x_2)}\mathbb{C} [x_1,x_2,x_3]
\end{eqnarray}
produces a sheaf associated with a D2-brane along $x_1=x_2=0$ and similarly for D2-branes of other orientations and their intersection. 

We have already found a coherent sheaf modeling an arbitrary configuration of branes specified by an assignment of a multiplicity to each cycle from the table \ref{branes}. But the world of quasi-coherent sheaves is much richer. For example, one can easily see that the support of
\begin{eqnarray}
\mathbb{C}[x_1,x_2,x_3]/(x_3^2)
\end{eqnarray}
agrees with the support of a D4-brane along $x_3=0$ but the module structure for the structure sheaf is obviously different. As a module for $\mathbb{C}[x_1,x_2]$, it is isomorphic to a direct sum of two D4-branes but the action of $x_3$ is now modified. We can think about such a sheaf as a deformation of a pair of D4-branes by turning on a nilpotent vacuum expectation value for the Higgs field living on their support \cite{2004nn}. As we are going to argue later in these notes, such a modification has a very nice consequence in the BPS-algebras world. 

We have just established a correspondence between branes and coherent sheaves. We have also seen that some of the sheaves we defined are somehow equivalent to complexes of different sheaves. Let us thus extend our category of coherent sheaves by allowing more general complexes and identify precisely which complexes describe equivalent configurations. This transition to complexes has also a physical interpretation. A general complex can be thought of as describing a non-trivial bound state of branes and anti-branes. The equivalence of complexes then accounts for processes known as tachyon condensation. See footnotes bellow for more details regarding this physical interpretation. 

Let us thus extend the category of coherent sheaves to admit complexes of sheaves of the form 
\begin{eqnarray}
\xymatrixcolsep{3pc}
\xymatrixrowsep{3pc}
\xymatrix{  \dots \ar[r]^{d_0} &  A_1\ar[r]^{d_1}& A_2 \ar[r]^{d_2} & A_3 \ar[r]^{d_3}  &\dots }
\end{eqnarray}
with the differential squaring to zero $d_{i+1}\circ d_i=0$. In this complex, $A_i$ are sheaves describing the branes from the above discussion and differentials $d_i$ specify the exact form of a bound state. 

The above complexes of coherent sheaves form elements of the derived category of coherent sheaves (our brane category). But introducing complexes, one needs to be careful with distinguishing physically non-equivalent objects. In order to do that, let me first define a quasi-isomorphism as a chain morphism
\begin{eqnarray}
\xymatrixcolsep{3pc}
\xymatrixrowsep{3pc}
\xymatrix{ \dots \ar[r]^{d_0}   &  A_1\ar[r]^{d_1} \ar[d]^{f_{1}} &  A_2\ar[r]^{d_1}  \ar[d]^{f_{2}}& A_3 \ar[r]^{d_2} \ar[d]^{f_{3}} &\dots \\
\dots \ar[r]^{d'_0} &  B_1\ar[r]^{d'_1}& B_2 \ar[r]^{d'_2} & B_3 \ar[r]^{d'_3}   &\dots &}
\label{chainmap}
\end{eqnarray}
satisfying $f_{i+1}\circ d_i=d'_i\circ f_i$ and inducing isomorphism on the cohomology. Two chain complexes $A$ and $B$ are then isomorphic in the derived category of coherent sheaves if there exists a third object $C$ with quasi-isomorphisms $f:C\rightarrow A$ and $g:C\rightarrow B$. Such $A$ and $B$ then describe an equivalent\footnote{This equivalence can be given a physical interpretation as well. Brane-anti-brane systems admit an unstable tachyonic string mode. In our story, you should be thinking of even-degree sheaves $A_{2n}$ as branes and odd-degree sheaves $A_{2n+1}$ as anti-branes. In the process of tachyon condensation \cite{Sen:1998sm}, the tachyonic field develops a non-trivial vacuum expectation value encoded in maps $d_i$. The endpoint of such a tachyon condensation describes a new configuration of branes related to the previous one by quasi-isomorphism.} physical configuration.   

Let me give two examples of a quasi-isomorphism and equivalent objects. First, we have an obvious exact\footnote{It is straightforward to check the exactness. The kernel of the first map $x_3$ is obviously zero. The kernel of $d$ are functions that vanish at $x_3=0$, i.e. functions of the form $x_3f(x_1,x_2,x_3)$ but these agree with the image of $x_3$. Finally, $d$ obviously generates whole $\mathbb{C}[x_1,x_2]$.} sequence from the above
\begin{eqnarray}
\xymatrixcolsep{3pc}
\xymatrixrowsep{3pc}
\xymatrix{  0 \ar[r] & \mathbb{C}[x_1,x_2,x_3] \ar[r]^{x_3} &  \mathbb{C}[x_1,x_2,x_3] \ar[r]^{d} & \frac{\mathbb{C}[x_1,x_2,x_3]}{(x_3)} \ar[r] & 0  },
\label{exact}
\end{eqnarray}
where the map $d$ simply sets $x_3=0$ in any polynomial $f\in  \mathbb{C}[x_1,x_2,x_3]$. Let me bend the complex and extend by zeros as follows
\begin{eqnarray}
\xymatrixcolsep{3pc}
\xymatrixrowsep{3pc}
\xymatrix{  \mathbb{C}[x_1,x_2,x_3] \ar[r]^{x_3} \ar[d]&  \mathbb{C}[x_1,x_2,x_3]\ar[d]^{d}  \\
  0 \ar[r] &\frac{\mathbb{C}[x_1,x_2,x_3]}{(x_3)}}.
\label{bend}
\end{eqnarray}
The map $d$ obviously induces an isomorphism on the cohomology and we can see that the sheaf we identified with the D4-brane along $x_3=0$ is quasi-isomorphic to the complex
\begin{eqnarray}
\xymatrixcolsep{3pc}
\xymatrixrowsep{3pc}
\xymatrix{\mathbb{C}[x_1,x_2,x_3] \ar[r]^{x_3}&  \mathbb{C}[x_1,x_2,x_3]  }.
\end{eqnarray}
These two are thus isomorphic objects in the derived category of coherent sheaves.\footnote{Returning to the physical interpretation of this statement, note that both elements of the complex are associated with the space-filling D6-brane with each being placed at a different cohomological degree. The shift of the degree by one distinguishes a brane form an anti-brane. The full system can be thus viewed as a non-trivial space-filling brane-anti-brane bound state. The brane-anti-brane configuration is unstable due to the existence of a tachyonic mode in the string spectrum (a mode with negative energy). The map $x_3$ can be then thought of as determining the tachyonic profile and giving rise (via the process of tachyon condensation) to a single D4-brane along $x_3=0$.} The resolution of lower-dimensional branes as bound states of space-filling branes will play an important role in the calculation of morphisms in our category of branes (derived category of coherent sheaves) and the derivation of quivers.

Let us consider yet another exact sequence   
\begin{eqnarray}
\xymatrixcolsep{3pc}
\xymatrixrowsep{3pc}
\xymatrix{ 0 \ar[r] &   \mbox{Ker}\ d  \ar[r]  & \mathbb{C}[x_1,x_2,x_3] \ar[r]^{d} &  \frac{\mathbb{C}[x_1,x_2,x_3]}{(x_1,x_2,x_3)} \ar[r] & 0  }
\label{ex2}
\end{eqnarray}
where the kernel of $d$ is simply generated by elements vanishing at the origin
\begin{eqnarray}
x_1f_1(x_1,x_2,x_3)+x_2f_2(x_1,x_2,x_3)+x_3f_3(x_1,x_2,x_3)
\end{eqnarray}
with the obvious embedding map to $\mathbb{C}[x_1,x_2,x_3]$. Such a sheaf is isomorphic to $\mathbb{C}[x_1,x_2,x_3]$ at a generic point but it caries a non-trivial modification at the origin. Bending the complex, we find a quasi-isomorphism
\begin{eqnarray}
\xymatrixcolsep{3pc}
\xymatrixrowsep{3pc}
\xymatrix{ \mbox{Ker}\ d  \ar[r]  \ar[d]&  0\ar[d] \\
  \mathbb{C}[x_1,x_2,x_3]  \ar[r]^{d} &\frac{\mathbb{C}[x_1,x_2,x_3]}{(x_1,x_2,x_3)}}.
\end{eqnarray}
 We can interpret the sheaf $\mbox{Ker}\ d $ as describing a non-trivial bound state of a D6-brane with a D0-brane. Bound states of this form together with their quiver description will be the main object of interest in our discussion. The form of this complex also justifies the notation $A\rightarrow n\ \mbox{D0}$ with $A$ being a system of non-compact branes from the introduction (in the present example $A=$D6).

\subsection{Morphisms in the brane category}

Our ultimate goal is understanding the low-energy dynamics of D0-branes (or more generally compactly supported branes if we go beyond the  $\mathbb{C}^3$ example) bound to a system of non-compact branes. Such a dynamics should have an effective description in terms of a field theory living on the support of the D0-branes. To find such a description, one usually studies the spectrum of open strings in a given background of D-branes \cite{Witten:1994tz,Douglas:1995bn,Douglas:1996sw,Nekrasov:2016gud} and identifies the physical masless modes that give rise to fields of the effective quantum mechanics. 

We would like to now understand how to identify physical massless modes of strings stretched between branes $A$ and $B$ in terms of morphisms $\mbox{Hom}^n(A,B)$ in the derived category of coherent sheaves. In any good enough\footnote{A derived category is good enough if it contains enough projective objects such that any object in the category is quasi-isomorphic to a complex of projective objects.} derived category, one can identify $\mbox{Hom}^n(A,B)$ with standard morphisms in the homotopy category (denoted as $\mbox{Hom}^n_{K}(\tilde{A},\tilde{B})$) between projective resolutions\footnote{It is actually enough for only $\tilde{A}$ to be a projective resolution or alternatively $\tilde{B}$ to be an injective resolution.} $\tilde{A},\tilde{B}$ of branes $A,B$. Let us start with decodifying this statement.

The starting point in the calculation of $\mbox{Hom}(A,B)$ is a projective resolution of $A,B$, i.e. an exact sequence of the form
\begin{eqnarray}
\xymatrixcolsep{3pc}
\xymatrixrowsep{3pc}
\xymatrix{  \dots \ar[r]^{d_{-4}} &  A_{-3}\ar[r]^{d_{-3}}& A_{-2} \ar[r]^{d_{-2}} & A_{-1} \ar[r]^{d_{-1}}  & A}
\end{eqnarray}
with all $A_i$ being projective.\footnote{In a usual situation, it it is often difficult to find a projective resolution explicitly. It is much easier to find a resolution in terms of locally-free sheaves associated with holomorphic bundles. The extension groups can be then determined using a so-called  local-to-global spectral sequence. Fortunately, this machinery is not necessary in the problem at hand and we recommend interested reader to consult the literature for further details.} In our examples, we will be able to find a resolution of all the relevant sheaves in terms of free sheaves of the form $\mathbb{C}[x_1,x_2,x_3]^{k}$ that are automatically projective. 

Let us now find projective resolutions of the elementary sheaves from the previous section (where we label $\mathcal{O}=\mathbb{C}[x_1,x_2,x_3]$ for simplicity):\footnote{It is straightforward to check the exactness along the lines of footnote checking the exactness of \ref{exact}. Note that we could also extend the complexes on the left by zeros.}
\begin{itemize}
\item The projective resolution of a D0-brane at the origin is given by
\begin{eqnarray}
\mathcal{O}\xrightarrow{\scriptsize{\begin{pmatrix}-x_1\\ x_2\\ -x_3 \end{pmatrix}}}\mathcal{O}^3\xrightarrow{\scriptsize{\begin{pmatrix}0&-x_3&-x_2\\ -x_3&0&x_1\\ x_2& x_1& 0 \end{pmatrix}}}\mathcal{O}^3\xrightarrow{\scriptsize{\begin{pmatrix}x_1&x_2&x_3 \end{pmatrix}}}\mathcal{O}\rightarrow \frac{\mathcal{O}}{(x_1,x_2,x_3)}.
\label{d0}
\end{eqnarray}
\item The projective resolution associated with the D2-brane supported along the coordinate line $x_1=x_2=0$ is
\begin{eqnarray}
\mathcal{O}\xrightarrow{\scriptsize{\begin{pmatrix}-x_2\\ x_1\end{pmatrix}}}\mathcal{O}^2\xrightarrow{\scriptsize{\begin{pmatrix}x_1&x_2 \end{pmatrix}}}\mathcal{O}\rightarrow \frac{\mathcal{O}}{(x_1,x_2)}
\end{eqnarray}
and analogously for D2-branes along $x_1=x_3=0$ and $x_2=x_3=0$. 
\item  The projective resolution associated with the D4-brane supported at plane $x_3=0$ is
\begin{eqnarray}
\mathcal{O}\xrightarrow{\scriptsize{x_3}}\mathcal{O}\rightarrow \frac{\mathcal{O}}{(x_3)}
\end{eqnarray}
and analogously for D4-branes along $x_2=0$ and $x_3=0$.
\item Finally, the space-filling D6-brane is represented by $\mathcal{O}$ itself
\begin{eqnarray}
\mathcal{O}\rightarrow \mathcal{O}.
\end{eqnarray}
\end{itemize}
As discussed above, $\mathcal{O}^k$ are projective objects and it is easy to check exactness of the complexes above. Bending the complexes at the last arrow analogously to the above discussion of D4-branes in (\ref{bend}), we can see that the rightmost sheaves are quasi-isomorphic to the complexes of projective objects obtained by stripping off the rightmost factor. We are now going to use such complexes to compute $\mbox{Hom}^n(A,B)$

The space of morphisms $\mbox{Hom}^n(A,B)$ can be identified with chain maps between the two projective resolutions modulo chain homotopies. Let us spell out explicitly what a chain map and a chain homotopy is. Let
\begin{eqnarray}\nonumber
\xymatrix{  \dots \ar[r]^{d_{-4}} &  A_{-3}\ar[r]^{d_{-3}}& A_{-2} \ar[r]^{d_{-2}} & A_{-1} },\\
\xymatrix{  \dots \ar[r]^{d'_{-4}} &  B_{-3}\ar[r]^{d'_{-3}}& B_{-2} \ar[r]^{d'_{-2}} & B_{-1}  }
\end{eqnarray}
be projective resolutions of $A$ and $B$. For a fixed $n$ and any $m<0$, let $f_{n,m}:A_m\rightarrow B_{m+n}$ be a collection of maps between the entries of the two complexes. For example, the collection $\{ f_{1,n}\}$ can be encoded in a diagram
\begin{eqnarray}
\xymatrixcolsep{3pc}
\xymatrixrowsep{3pc}
\xymatrix{ \dots \ar[r]^{d_{-5}}   &  A_{-4}\ar[r]^{d_{-4}} \ar[d]^{f_{1,-4}} &  A_{-3}\ar[r]^{d_{-3}}  \ar[d]^{f_{1,-3}}& A_{-2} \ar[r]^{d_{-2}} \ar[d]^{f_{1,-2}} & A_{-1}  \\
\dots \ar[r]^{d_{-4}}& B_{-3} \ar[r]^{d_{-3}} & B_{-2} \ar[r]^{d_{-2}}  & B_{-1}   & }
\label{chainmap}
\end{eqnarray}
where we shifted the second complex by $n=1$ to the left. 

Let us now define a differential $\partial:f_{n,m}\rightarrow f_{n+1,m}$ acting on a collection of map $\{ f_{n,m}\}$ and producing a collection $\{ f_{n+1,m}\}$. The differential $\partial$ is defined by formula 
\begin{eqnarray}
\partial f_{n,m}=d_{m+n}\circ f_{n,m}-(-1)^{n}f_{n,m+1}\circ d_m.
\end{eqnarray}
A collection of $f_{n,m}$ for fixed $n$ is called a chain map if it lies in the kernel of this map (this is equivalent to all the squares in (\ref{chainmap}) commuting or anti-commuting). Chain homotopies then correspond to the image of $\partial$. The spectrum of strings $\mbox{Hom}^n(A,B)$ can be thus identified with the cohomology of $\partial$ acting on the collection of maps $f_{n,m}$ for $m<0$. The integer $n$ is called the ghost number. The subsector of degree-one maps $\mbox{Hom}^1(A,B)$ is sometimes called the spectrum of physical-string modes. This space will be of our main interest in the derivation of our quiver quantum mechanics.

\subsection{Framed quivers}

In the rest of the chapter, we are going to adapt the above tools to derive framed quivers with potential encoding the spectrum of physical massless string modes modes and in turn describing the low-energy dynamics of D0-branes bound to a fixed configuration of non-compactly supported branes. 

As discussed in the introduction, the low-energy dynamics of compactly supported D-branes bound to a general configuration non-compact branes should be captured by an $\mathcal{N}=4$ sypersymmetric gauged quantum mechanics with potential (sometimes called superpotential). Without diving too much into details of their construction, such quantum mechanics is uniquely specified by the following data:
\begin{enumerate}
\item A gauge group $G$ specifying fields forming so-called vector multiplets.
\item A representation $M$ of the gauge group $G$ specifying fields forming so-called chiral multiplets.
\item A $G$-invariant holomorphic function on $M$ called the potential $W$.
\end{enumerate}
The details of the construction of such a supersymmetric quantum mechanics are not going to be important for our purposes and they can be found e.g. in \cite{Ohta:2014ria}. The purpose of this section is a derivation of all of this data from calculations in the derived category of coherent sheaves. We are going to closely follow \cite{Aspinwall:2004bs} and extend their discussion by adding framing by non-compactly-supported branes \cite{BR}.

The pair $(G,M)$ arising from the compactification of a system of branes in toric Calabi-Yau threefolds can be encoded in terms of a quiver diagram (see e.g. \cite{yan2014a} for a nice review). The gauge group $G$ is generally a product of $U(n_i)$ factors. Each factor is associated with a generator of the subcategory of compactly-supported branes. We associate a circular node labeled by integer $n_i$ with  each such $U(n_i)$ factor. The integer $n_i$ denotes the number of corresponding compactly-supported branes in the system at hand. Since all the compactly supported branes in our $\mathbb{C}^3$ example are D0-branes, we have a single node with label $n$ specifying the number of such D0-branes:
\[
\begin{tikzcd}
\mathcircled{n}
\end{tikzcd}
\]

Analogously, we associate a square (framing) node with each possible elementary non-compactly supported brane in the given geometry. We assign a multiplicity $k_j$ to each such node, determining the number\footnote{We define the rank as a number of generators of the module for the structure sheaf of the subvariety on which the brane is supported. For example $\mathbb{C}[x_1,x_2,x_3]/(x_3^2)$ would be of rank two since it can be identified with $\mathbb{C}[x_1,x_2]+\mathbb{C}[x_1,x_2]x_3$ as a module for $\mathbb{C}[x_1,x_2]$. Note that one needs to be careful when identifying the rank for intersecting branes. The ranks are associated to each smooth component of the subvariety and computed away of the intersection.} of elementary branes of a given orientation. The system of a $k$ D4-brane along $\mathbb{C}_{\epsilon_1}\times \mathbb{C}_{\epsilon_2}$ bound to $n$ D0-branes would lead to:
\[
\begin{tikzcd}
\boxed{k} & \mathcircled{n}
\end{tikzcd}
\]

The representation $M$ of $G$ is then specified by a set of arrows between nodes in the diagram, $M=\oplus_{a}M_a$ with the sum running over all arrows. Each factor $M_a$ associated with an arrow $a$ starting at vertex $t(a)$ and ending at vertex $h(a)$ can be identified with a morphism $M_a\in \mbox{Hom}(\mathbb{C}^{n_{t(a)}},\mathbb{C}^{n_{h(a)}})$. An element $m_a\in M_a$ then transforms in the fundamental representation of the $U(n_i)$ factor and in the anti-fundamental representation of $U(n_i)$, i.e.
\begin{eqnarray}
g_1\times \dots \times g_m:m_i\rightarrow g_{h(i)}m_i g_{t(i)}^{-1}
\end{eqnarray}
for $g_1\times \dots \times g_n \in G= U(n_1)\times \dots \times U(n_m)$. This in particular means that $M_i$ associated with loops transforms in the adjoin representation. Note also that only the circular nodes contribute to $G$ (the dynamics of non-compactly supported branes is frozen). So, an arrow going from a circular node $t(i)$ to a square note $h(i)$ would contribute by $k_{h(i)}$ copies of the anti-fundamental representations of the factor $U(n_{t(i)})$ and analogously for an arrow of the opposite orientation. For example, we are going to see that the diagram for $k$ D4-brane along $\mathbb{C}_{\epsilon_1}\times \mathbb{C}_{\epsilon_2}$ bound to $n$ D0-branes is going to be
\[
\begin{tikzcd}
\boxed{k} \arrow[r,   shift left=0.5ex, "I"] & \arrow[l, shift left=0.5ex,   "J"]  \mathcircled{n}\arrow[out=340,in=20,loop,swap,"B_3"]
  \arrow[out=70,in=110,loop,swap,"B_2"]
  \arrow[out=250,in=290,loop,swap,"\normalsize{B_1}"]
\end{tikzcd}
\]
The gauge group is thus $U(n)$, the matter content\footnote{Note that we label arrows attached to the circular nodes on both sides by letter $B$, arrows going from the square node to the circular node by $I$ and arrows going in the opposite direction by $J$. We are going to implement this convention throughout the note. Later, we are also going to introduce arrows joining framing nodes and assign label $A$ to them.} consists of $M=B_1\oplus B_2 \oplus B_3 \oplus I\oplus J$, where $B_i=\mbox{Hom}(\mathbb{C}^n, \mathbb{C}^n)$, $I=\mbox{Hom}(\mathbb{C}^k, \mathbb{C}^n)$ and $J=\mbox{Hom}(\mathbb{C}^n, \mathbb{C}^k)$ and they are acted on by $g\in U(n)$ as
\begin{eqnarray}
g: b_i \rightarrow gb_ig^{-1},\qquad g:i\rightarrow ig,\qquad g:j\rightarrow jg
\end{eqnarray}
for any $b_i\in B_i,i\in I,j\in J$.

Physically, fields associated with arrows joining two circular nodes should arise from massless physical modes of strings stretched between two compactly supported branes and fields associated with arrows joining a circular node with a square node should arise from   massless physical modes of strings stretched between a compactly supported brane and a brane of non-compact support. Instead of a cumbersome analysis of the physical string (see e.g. \cite{Nekrasov:2016gud}), we are going employ the language of derived categories of coherent sheaves. As already discussed in the previous sections, different systems of branes correspond to objects in the derived category of coherent sheaves. Massless physical strings stretched between them correspond to morphisms in such a category. As a result, arrows in our quiver diagrams will be identified with generators of $\mbox{Hom}^1(A,B)$, where at least one of $A,B$ is a sheaf associated with a compactly-supported brane. 

Note that the appearance of $\mathbb{C}^{n_{t(i)}}$ and $\mathbb{C}^{n_{h(i)}}$ in $M_i=\mbox{Hom}(\mathbb{C}^{n_{t(i)}}, \mathbb{C}^{n_{h(i)}})$ has also a physical explanation.  Introducing multiple branes of a given elementary support introduces a degeneracy in the space of string modes. These need to be parametrized by two indices, each labeling the elementary brane on which the string ends on each side leading us to the factors $\mathbb{C}^{n_{t(i)}}$ and $\mathbb{C}^{n_{h(i)}}$.

One needs to be careful about the relative shift\footnote{Note that the sheaf associated with D6-brane is one entry on the left from the D0-brane sheaf in the example (\ref{ex2})} of objects we want to investigate. Remember that we are interested in the system of D0-branes bound to higher-dimensional branes. As explained above, construction of a bound states requires the degree of one of the two branes to be shifted. We thus need to introduce shifts of complexes associated with non-compact branes. 

A complex $A$ admits natural shifts by an integer $n$ producing a new complex $A[n]$ with entries $\tilde{A}_m=A[n]_m=A_{m+n}$ and differentials $\tilde{d}_m=d_{m+n}$. This operation simply shifts the relevant complex by $n$ steps to the left. To obtain the desired quivers we need to shift the degree of non-compactly supported branes by $1$.

It is now straightforward to determine the dimensions of $\mbox{Hom}^n(A,B)$ involving D0-branes, producing
\begin{center}
\begin{tabular}{| c | c c  c c|}\hline
  $\mbox{dim}\ \mbox{Hom}^n$ & $n=0$ & $n=1$& $n=2$& $n=3$\\ \hline
D0-D0 &  1 & 3  & 3  & 1  \\ \hline
D0-D6{[}1{]} &   &   &  1 &   \\
D6{[}1{]}-D0 &   & 1  &   &   \\ \hline
D0-D4{[}1{]} &   &  1 & 1  &   \\
D4{[}1{]}-D0 &   & 1  & 1  &   \\ \hline
D0-D2{[}1{]} &  1 & 2  & 1  &   \\ 
D2{[}1{]}-D0 &   & 1  & 2  & 1 \\ \hline
\end{tabular}
\label{tab}
\end{center}
We are going to see explicit examples of $\mbox{Hom}^n(A,B)$ generators recovering these numbers in the next section.

Using the above table, it is easy to read off the quiver associated with a given system of non-compact branes by:
\begin{enumerate}
\item drawing a circular node associated with D0-branes with an integer $n$ indicating their number,
\item drawing a square node for each elementary non-compact brane and associating with them numbers $k_i$ indicating their multiplicity,
\item attaching a loop to the circular node associated with each generator of $\mbox{Hom}^1(D0,D0)$
\item attaching an arrow from the circular node to one of the square nodes for each generator of $\mbox{Hom}^1(\mbox{D0},A)$ and
\item attaching an arrow from one of the square nodes to the circular node for each generator of $\mbox{Hom}^1(A,\mbox{D0})$ 
\end{enumerate}

\subsection{Higgs field}

Above, we have described how to identify quivers associated with direct sums of elementary non-compactly supported branes. Let us now discuss the modification necessary for describing D0-branes bound to more general sheaves, such as $\mathbb{C}[x_1,x_2,x_3]/(x_3)$. As already mentioned in the introduction, these should correspond to turning on a non-trivial vacuum expectation value for the Higgs field on the non-compact brane \cite{2004nn}. Such a Higgs field is again in correspondence with generators of $\mbox{Hom}^1(A,B)$ but now between non-compact branes.  Similarly as in the calculation above, we can now identify\footnote{Note that the result is independent of the shift of both branes simultaneously $\mbox{Hom}^n(A,B)=\mbox{Hom}^n(A[1],B[1])$ and we have thus suppressed the shifts for the purposes of this table.}
\begin{center}
\begin{tabular}{| c | c c  c c|}\hline
  $\mbox{dim  Hom}^n$ & $n=0$ & $n=1$& $n=2$& $n=3$\\ \hline
D6-D6 &  1 &   &   &   \\ \hline
D4${}_{12}$-D4${}_{12}$ & 1  & 1  &   &   \\ 
D4${}_{12}$-D4${}_{13}$ &   &  1 &   &   \\  \hline
D2${}_{1}$-D2${}_{1}$ & 1  & 2  & 1  &  \\  
D2${}_{1}$-D2${}_{2}$ &   & 1  & 1  &  \\  \hline
\end{tabular}
\end{center}
Note that the table shows only the rank of $\mbox{Hom}^n(A,B)$ as a module for the ring of functions at the brane intersection. For example, the full $\mbox{Hom}^1(\mbox{D}4_{13},\mbox{D}4_{12})$ would be isomorphic to $\mathbb{C}[x_1]$ since the two branes intersect along $\mathbb{C}_{\epsilon_1}$. In this calculation, we also need to keep track of the mutual orientation of involved branes. It is straightforward to determine the morphism spaces also between branes of different dimensions and I leave it as an exercise to the reader.

It is tempting to attach extra loops to the framing nodes and arrows between them according to the table above. But we have to be careful since these modes are non-dynamical from our perspective (they parametrize the motion of non-compact branes) and should be treated differently. To make this distinction clear, we are going to draw a point in the middle of such an added arrow. Though non-dynamical, we can still turn on a non-trivial constant vacuum expectation value for the Higgs field by specifying the precise form of the maps (modulo change of the basis). 

As we are going to see later, generic vacuum expectation values for the Higgs field would be inconsistent with the $U(1)^2$ invariance required by the $\Omega$-background. This fact is in agreement with the interpretation that eigenvalues of the Higgs field parametrize the motion of branes in orthogonal directions. On the other hand, turning on a nilpotent vacuum expectation value does not spoil such an invariance and is a legitimate deformation of our system.

We are now going to see how turning on a nilpotent vacuum expectation value for the Higgs field allow us to study more general class of sheaves. Let us illustrate the construction in two examples. First, let us show how to realize the coherent sheave $\mathbb{C}[x_1,x_2,x_3]/(x_3^2)$ in terms of a pair of sheaves associated with the D4-brane along $\mathbb{C}_{\epsilon_3}$. We can read off from the above table that $\mbox{Hom}^1(D4,D4)$ is of dimension one and let us label the corresponding Higgs field by $A$. Since $\mathbb{C}[x_1,x_2,x_3]/(x_3^2)$ is of dimension two as a $\mathbb{C}[x_1,x_2]$-module, it should arise from a deformation of a stack of two D4-branes. The Higgs field is thus a $2\times 2$ matrix and let us give it a vacuum expectation value
\begin{eqnarray}
A=\begin{pmatrix}
0&0\\
1&0
\end{pmatrix}.
\end{eqnarray} 
We can now modify the action of $x_3$ on 
\begin{eqnarray}
P=  \begin{pmatrix}P_1(x_1,x_2)& P_2(x_1,x_2)\end{pmatrix} \in \mathbb{C}^2\otimes \frac{\mathbb{C}[x_1,x_2,x_3]}{(x_3)}
\label{pcko}
\end{eqnarray}
by
\begin{eqnarray}
x_3 P=PA.
\label{twist}
\end{eqnarray}
The resulting module is obviously isomorphic to the $\mathbb{C}[x_1,x_2,x_3]/(x_3^2)$ module if we identify its elements with $P_2(x_1,x_2)+x_3 P_1(x_1,x_2)\in \mathbb{C}[x_1,x_2,x_3]/(x_3^2)$. We are going to see soon that the conditions of the form (\ref{twist}) can be naturally derived from the variation of the potential. Similarly, considering $(\mathbb{C}[x_1,x_2,x_3]/(x_3))^n$ and turning on the nilpotent Higgs-field vacuum expectation value with $1$ above the diagonal produces a module isomorphic to $\mathbb{C}[x_1,x_2,x_3]/(x_3^n)$. An analogous procedure leads to more general bound states associated with $\mathbb{C}[x_1,x_2,x_3]/(x_1^{N_1}x_2^{N_2}x_3^{N_3})$.

One can similarly proceed with an analysis of non-trivial bound states of D2-branes. Let us again start with a simple example. The dimension of $\mbox{Hom}^1(D2,D2)$ is two and we now have two Higgs fields $A_1,A_2$. In order to realize the module $\mathbb{C}[x_1,x_2,x_3]/(x_1,x_2^2)$ in terms of a pair of D2-branes along $\mathbb{C}_{\epsilon_1}$, one needs to introduce a pair of matrices of the form
\begin{eqnarray}
A_1=\begin{pmatrix}
0&0\\
0&0
\end{pmatrix}\qquad 
A_2=\begin{pmatrix}
0&0\\
1&0
\end{pmatrix}
\end{eqnarray} 
acting on
\begin{eqnarray}
P=  \begin{pmatrix}P_1(x_3)& P_2(x_3)\end{pmatrix} \in \mathbb{C}^2\otimes \frac{\mathbb{C}[x_1,x_2,x_3]}{(x_1,x_2)}
\label{pcko}
\end{eqnarray}
and modify the action of $x_1$ and $x_2$ by
\begin{eqnarray}
x_1P=PA_1,\qquad x_2P=PA_2.
\label{twist2}
\end{eqnarray}
It is easy to see that identifying $P_2(x_1)+x_2 P_1(x_3)\in\mathbb{C}[x_1,x_2,x_3]/(x_1,x_2^2) $ gives an isomorphism of the above $\mathbb{C}[x_1,x_2,x_3]$-module with $\mathbb{C}[x_1,x_2,x_3]/(x_1,x_2^2)$. 

More generally, we can associate a sheaf with the support along $\mathbb{C}_{\epsilon_3}$ with any 2d partition. Concretely, the partition encodes the nilpotency structure of $A_1,A_2$ and modifying the action of $x_1$ and $x_2$ on $(\mathbb{C}[x,y,z]/(x_1,x_3))^n$ as above produces the desired sheaf. 

\begin{figure}
  \centering
    \includegraphics[width=0.14\textwidth]{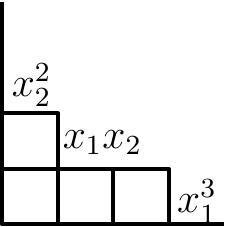}
\caption{The corners carving out the partition are at positions $(m,n)$ for generators of the ideal $x_1^mx_2^n$. In our example the ideal is $(x_1^3,x_1x_2,x_2^2)$. The quotient module is then generated by $1,x_1,x_1^2,x_2$ corresponding to the boxes in the diagram.}
\label{partition}
\end{figure}

Let us give one more example of the correspondence between a partition and a $\mathbb{C}[x_1,x_2,x_3]$-module by looking at $\mathbb{C}[x_1,x_2]/(x_1^3,x_1x_2,x_2^2)$. Generators of the module $\mathbb{C}[x_1,x_2]/(x_1^3,x_1x_2,x_2^2)$ are in correspondence with boxes of the Young diagram from figure \ref{partition} carved out by the generators of the ideal $x_1^3,x_1x_2,x_2^2$ In particular, the module is generated by $1,x_1,x_1^2,x_2$ with the powers in correspondence with coordinates of the boxes. We can now redraw the Young diagram in terms of a commutative diagram of the following form 
\begin{eqnarray}
\xymatrixcolsep{3pc}
\xymatrixrowsep{3pc}
\xymatrix{  0\ar[r]^{A_1} &  0\ar[r]^{A_1}&  0\ar[r]^{A_1}  &0  \\
      \mathbb{C}\ar[r]^{A_1} \ar[u]^{A_2}&0\ar[r]^{A_1} \ar[u]^{A_2}&0 \ar[r]^{A_1} \ar[u]^{A_2}&0 \ar[u]^{A_2}  \\
\mathbb{C}\ar[r]^{A_1} \ar[u]^{A_2} &  \mathbb{C} \ar[r]^{A_1} \ar[u]^{A_2}&  \mathbb{C}\ar[r]^{A_1} \ar[u]^{A_2} &0  \ar[u]^{A_2}}
\end{eqnarray}
with each box mapping to $\mathbb{C}$ and the horisontal lines associated with $A_1$ and the vertical lines associated with $A_2$. It is easy to check that that we can fix a basis of $\mathbb{C}^4$ such that matrices 
\begin{eqnarray}
A_1=\begin{pmatrix}
0 & 0 &0 &0 \\
1 & 0 &0 &0 \\
0 & 1 &0 &0 \\
0 & 0 &0 &0 \\
\end{pmatrix},\qquad 
A_2=\begin{pmatrix}
0 & 0 &0 &0 \\
0 & 0 &0 &0 \\
0 & 0 &0 &0 \\
1 & 0 &0 &0 \\
\end{pmatrix}
\end{eqnarray}
decompose $\mathbb{C}^4$  exactly as indicated in the diagram. One can proceed analogously with more complicated partitions.

\subsection{Potential}

Our last step is a determination of the potential $W$. We have already identified the massless string modes with generators of $\mbox{Hom}^*(A,B)$. Strings can also mutually join and split leading to an associative product (usually called the star product in the string field theory literature)
\begin{eqnarray}
m_2: \mbox{Hom}^*(A_1,A_2) \otimes \mbox{Hom}^*(A_2,A_3) \rightarrow \mbox{Hom}^*(A_1,A_3).
\end{eqnarray} 
Note that the branes $A_2$ associated with the endpoints we want to merge need to agree for the product to make sense. The star product $m_2$ is a member of an infinite family of higher products
\begin{eqnarray}
m_n: \mbox{Hom}^*(A_1,A_2) \otimes \mbox{Hom}^*(A_2,A_3) \otimes \dots \otimes \mbox{Hom}^*(A_n,A_{n+1}) \rightarrow \mbox{Hom}^*(A_1,A_{n+1})
\end{eqnarray}
for any $n\geq 1$. The system of higher products $m_n$ satisfies a system of compatibility conditions and forms so-called $A_{\infty}$-structure in our brane category \cite{articlee,article,2001ff,2001,hha,Aspinwall:2004bs,2004mm,2005}.

A general procedure for deriving the $A_{\infty}$-structure is reviewed in \cite{Aspinwall:2004bs} and goes beyond the scope of these lecture notes. Using the general proposal, one can easily show that in our simplest example of $\mathbb{C}^3$, all $m_n=0$ vanish for any $n>2$. Furthermore, the only non-vanishing product $m_2$ is simply given by the composition of chain maps. For the purpose of this section, we are going to denote it as $m_2(\alpha,\beta)=\alpha \star \beta $.

Note also the symmetry of the first table from section \ref{tab} of the form
\begin{eqnarray}
\mbox{dim}\ \mbox{Hom}^n(\mbox{D}0,A)=\mbox{dim}\ \mbox{Hom}^{3-n}(A,\mbox{D}0)
\end{eqnarray}
for any $A=\mbox{D}0,\mbox{D}2[1],\mbox{D}4[1],\mbox{D}6[1]$. This observation is a consequence of the Serre duality for coherent sheaves in Calabi-Yau threefolds stating that there exists a natural pairing roughly of the form\footnote{A more general statement holds for varieties of any dimension and involves tensoring with a so-called dualizing sheaf. Luckily, the situation is much more trivial in our present example of $\mathbb{C}^3$.}
\begin{eqnarray}
\mbox{Hom}^n(D0,A)\times \mbox{Hom}^{3-n}(A,D0)\rightarrow \mathbb{C}.
\end{eqnarray}
This pairing can be written as 
\begin{eqnarray}
\int \alpha \star \beta
\end{eqnarray}
where $\int$ is known as the trace map. Let us identify a concrete form of the trace map in our example. First, note that $\mbox{Hom}^3(\mbox{D}0,\mbox{D}0)$ is generated by a single element that can be identified with the chain map
\begin{eqnarray}
\xymatrixcolsep{3pc}
\xymatrixrowsep{3pc}
\xymatrix{ & & & \mathcal{O}\ar[d]^{\scriptsize{1}} \ar[r] & \mathcal{O}^3\ar[r] &  \mathcal{O}^3 \ar[r] & \mathcal{O}  \\
         \mathcal{O} \ar[r] & \mathcal{O}^3 \ar[r] &  \mathcal{O}^3 \ar[r] & \mathcal{O}& & &}.
\label{dual}
\end{eqnarray}
More generally, if we have $n$ D0-branes, $\mbox{Hom}^3(n\mbox{D}0,n\mbox{D}0)$ can be identified with the very same element but now tensored with an $n\times n$ matrix. We then identify the trace map $\int$ on $\mbox{Hom}^3(n\mbox{D}0,n\mbox{D}0)$ with the standard trace of this $n\times n$ matrix.

Knowing the $A_{\infty}$-structure and the trace map, it is now easy to write down the potential. First, let us fix bases\footnote{We are going to find explicit representatives of these generators momentarily.} $\{b_k\}\in \mbox{Hom}^1(\mbox{D0},\mbox{D0})$, $\{i^{\alpha}_k\}\in \mbox{Hom}^1(A_\alpha,\mbox{D0})$ and  $\{j^\alpha_k\}\in \mbox{Hom}^1(\mbox{D0},A_\alpha)$ for each of the elementary branes $A_\alpha=D2_{1}[1],D2_2[1],D2_3[1],D4_{12}[1],D4_{13}[1],D4_{23}[1],D6[1]$. Fixing the number of non-compact branes $k_\alpha$ of of a given support $A_\alpha$ and the number of D0-branes, we can write down a general string mode of
\begin{eqnarray}
\mbox{Hom}^1(n\mbox{D0},n\mbox{D0})\oplus \mbox{Hom}^1(\oplus_{\alpha}k_\alpha A_\alpha,n\mbox{D0})\oplus \mbox{Hom}^1(n\mbox{D0},\oplus_{\alpha}k_\alpha A_\alpha)
\end{eqnarray}
as a linear combination
\begin{eqnarray}
\Psi= \sum_{k}B_kb_k+\sum_{\alpha,k}I^\alpha_ki^\alpha_k+\sum_{\alpha, k}J^\alpha_kj^\alpha_k,
\end{eqnarray} 
where  $B_k:\mathbb{C}^n\rightarrow \mathbb{C}^n$, $I_k^\alpha:\mathbb{C}^{k_\alpha}\rightarrow \mathbb{C}^n$, $J^\alpha_k:\mathbb{C}^n\rightarrow \mathbb{C}^{k_\alpha}$. The linear combination of string modes $\Psi$ is often called the string field. 

Turning on a non-trivial Higgs-field vacuum expectation value on $\oplus_{\alpha}k_\alpha A_{\alpha}$ requires an introduction of $\mbox{Hom}^1(\oplus_{\alpha}k_\alpha A_{\alpha},\oplus_{\alpha}k_\alpha A_{\alpha})$ and the string field modifies as
\begin{eqnarray}
\Psi\rightarrow \Psi +\sum_{\alpha,\beta,k}A^{\alpha\beta}_ka^{\alpha\beta}_k
\end{eqnarray} 
for $a^{\alpha\beta}_k$ generators of $\mbox{Hom}^1(A_\alpha,A_\beta)$ and $A^{\alpha\beta}_k:\mathbb{C}^{k_\alpha}\rightarrow \mathbb{C}^{k_\beta}$.

The general proposal for the potential as a function of $B_k,I^\alpha_k,J^\alpha_k$ is of the form
\begin{eqnarray}
W= \sum_{k=2}^{\infty}\frac{1}{k+1}\int \Psi \star  \mu_k(\Psi,\dots,\Psi) .
\end{eqnarray}
In the situation at hand, the higher products vanish and the potential reduces (up to the irrelevant normalization) to
\begin{eqnarray}
W=\int \Psi \star \Psi \star \Psi .
\end{eqnarray}
In the rest of this section, we are going to illustrated the above construction on a large family of examples.

\subsection{Examples of framed quivers with potential}

In this section, we are going to derive framed quivers with potential arising from different choices of the non-compact brane $A$ or their combinations.

\subsubsection{D0-D0 strings}

Let us start with an analysis of the part universal to all the framed quivers, i.e. the one arising from the D0-D0 strings. We have obviously the following three generators of $\mbox{Hom}^1(\mbox{D}0,\mbox{D}0)$:
\begin{eqnarray}\nonumber
b_1\quad :\quad 
\xymatrixcolsep{4pc}
\xymatrixrowsep{3pc}
\xymatrix{ & \mathcal{O}\ar[d]^{\scriptsize{\begin{pmatrix}1\\0\\0\end{pmatrix}}} \ar[r] & \mathcal{O}^3\ar[d]^{\scriptsize{\begin{pmatrix}0&0 & 0\\0&0&-1\\0&-1&0\end{pmatrix}}} \ar[r] &  \mathcal{O}^3\ar[d]^{\scriptsize{\begin{pmatrix}1,0,0\end{pmatrix}}} \ar[r] & \mathcal{O}  \\
         \mathcal{O} \ar[r] & \mathcal{O}^3 \ar[r] &  \mathcal{O}^3 \ar[r] & \mathcal{O}&}\\ \nonumber
\vspace{2cm}\\ 
b_2\quad :\quad
\xymatrixcolsep{4pc}
\xymatrixrowsep{3pc}
\xymatrix{ & \mathcal{O}\ar[d]^{\scriptsize{\begin{pmatrix}0\\-1\\0\end{pmatrix}}} \ar[r] & \mathcal{O}^3\ar[d]^{\scriptsize{\begin{pmatrix}0&0 & 1\\0&0&0\\-1&0&0\end{pmatrix}}} \ar[r] &  \mathcal{O}^3\ar[d]^{\scriptsize{\begin{pmatrix}0,1,0\end{pmatrix}}} \ar[r] & \mathcal{O}  \\
         \mathcal{O} \ar[r] & \mathcal{O}^3 \ar[r] &  \mathcal{O}^3 \ar[r] & \mathcal{O}&}\\ \nonumber
\vspace{2cm}\\ \nonumber
b_3\quad :\quad
\xymatrixcolsep{4pc}
\xymatrixrowsep{3pc}
\xymatrix{ & \mathcal{O}\ar[d]^{\scriptsize{\begin{pmatrix}0\\0\\1\end{pmatrix}}} \ar[r] & \mathcal{O}^3\ar[d]^{\scriptsize{\begin{pmatrix}0&1 & 0\\1&0&0\\0&0&0\end{pmatrix}}} \ar[r] &  \mathcal{O}^3\ar[d]^{\scriptsize{\begin{pmatrix}0,0,1\end{pmatrix}}} \ar[r] & \mathcal{O}  \\ \nonumber
         \mathcal{O} \ar[r] & \mathcal{O}^3 \ar[r] &  \mathcal{O}^3 \ar[r] & \mathcal{O}&}
\end{eqnarray} 
Remember that the vertical maps are given by the complex (\ref{d0}). It is straightforward to check the anti-commutativity of the maps and that they are not homotopic to trivial maps. Note also that the bottom complex is shifted by one to the left so that we get a generator of ghost number one (since we are dealing with maps of ghost-number one, the squares anti-commute). Finally, if we multiplied the maps by any generator of $\mathcal{O}$ other than a constant, the resulting chain map would actually be homotopy trivial. The chain maps $b_1,b_2,b_3$ thus generate full $Hom^1(\mbox{D}0,\mbox{D}0)$ and justify the dimension three from table \ref{tab}. For example, $b_1\star b_2$ produces
\begin{eqnarray}
\xymatrixcolsep{4pc}
\xymatrixrowsep{3pc}
\xymatrix{ & & \mathcal{O}\ar[d]^{\scriptsize{\begin{pmatrix}0\\0\\-1\end{pmatrix}}} \ar[r] & \mathcal{O}^3\ar[d]^{\scriptsize{\begin{pmatrix}0&0 & -1\end{pmatrix}}} \ar[r] &  \mathcal{O}^3 \ar[r] & \mathcal{O}  \\
         \mathcal{O} \ar[r] & \mathcal{O}^3 \ar[r] &  \mathcal{O}^3 \ar[r] & \mathcal{O}& &}
\end{eqnarray} 
and further composing with $b_3$ reproduces (\ref{dual}) and leads to the term $\mbox{Tr}\ B_1B_2B_3$ in the potential (up to sign) and similarly for other compositions.

Inserting our string field 
\begin{eqnarray}
\Psi=B_1b_1+B_2b_2+B_3b_3
\end{eqnarray}
into the formula for the potential, i.e. composing the corresponding maps, identifying the coefficient of the $\mbox{Hom}^3(\mbox{D}0,\mbox{D}0)$ generator from (\ref{dual}) and taking its trace, we get\footnote{Here and also everywhere bellow, we are not careful about the precise form of coefficients multiplying each of the terms. These can be absorbed into the normalization of the involved fields.}
\begin{eqnarray}
W=\mbox{Tr}\ B_1[B_2,B_3].
\label{pot1}
\end{eqnarray}

\subsubsection{D6-D0 system}

Introducing a single D6-brane, we need to add a single generator corresponding to the map:
\begin{eqnarray}
i\quad :\quad 
\xymatrixcolsep{3pc}
\xymatrixrowsep{3pc}
\xymatrix{ & & & \mathcal{O}\ar[d]^{\scriptsize{1}}   \\
         \mathcal{O} \ar[r] & \mathcal{O}^3 \ar[r] &  \mathcal{O}^3 \ar[r] & \mathcal{O}& & &}
\end{eqnarray}
Remember that non-compact branes are shifted by one to the left, offsetting the shift of the D0-brane in the calculation of ghost-number-one morphisms. Inserting 
\begin{eqnarray}
\Psi=B_1b_1+B_2b_2+B_3b_3+Ii
\end{eqnarray}
into the formula for the potential, we find it remains unchanged.

What we have recovered is the standard framed quiver of the form
\[
\begin{tikzcd}
\boxed{k} \arrow[r,   "I"] & \mathcircled{n}\arrow[out=340,in=20,loop,swap,"B_3"]
  \arrow[out=70,in=110,loop,swap,"B_2"]
  \arrow[out=250,in=290,loop,swap,"\normalsize{B_1}"]
\end{tikzcd}
\]
with potential (\ref{pot1}) from \cite{Moore:1998et,szendroi2016cohomological,Rapcak:2020ueh}.

\subsubsection{D0-D4 system}

Let us choose the orientation of a D4-brane to be along $\mathbb{C}_{\epsilon_2}\times \mathbb{C}_{\epsilon_3}$. The generators $b_1,b_2,b_3$ are of the same form as above but now we have the following generators of $\mbox{Hom}^1(\mbox{D}4[1],\mbox{D}0)$ and $\mbox{Hom}^1(\mbox{D}0,\mbox{D}4[1])$ respectively:
\begin{eqnarray}\nonumber
i\quad :\quad 
\xymatrixcolsep{3pc}
\xymatrixrowsep{3pc}
\xymatrix{ &  & \mathcal{O}\ar[d]^{\scriptsize{\begin{pmatrix}0\\0\\ 1\end{pmatrix}}} \ar[r] &  \mathcal{O}\ar[d]^{\scriptsize{1}} \\
         \mathcal{O} \ar[r] & \mathcal{O}^3 \ar[r] &  \mathcal{O}^3 \ar[r] & \mathcal{O}}\\
\vspace{2cm}\\  \nonumber
j\quad :\quad 
\xymatrixcolsep{3pc}
\xymatrixrowsep{3pc}
\xymatrix{ \mathcal{O} \ar[d]^{-1} \ar[r] & \mathcal{O}^3 \ar[d]^{\scriptsize{\begin{pmatrix}0&0&1 \end{pmatrix}}} \ar[r] &  \mathcal{O}^3\ar[r] & \mathcal{O}  \\
       \mathcal{O} \ar[r] & \mathcal{O}   && }
\end{eqnarray}
The quiver associated with the D4-D0 system is thus given by
\[
\begin{tikzcd}
\boxed{k} \arrow[r,   shift left=0.5ex, "I"] & \arrow[l, shift left=0.5ex,   "J"]  \mathcircled{n}\arrow[out=340,in=20,loop,swap,"B_3"]
  \arrow[out=70,in=110,loop,swap,"B_2"]
  \arrow[out=250,in=290,loop,swap,"\normalsize{B_1}"]
\end{tikzcd}
\]
Inserting the string field
\begin{eqnarray}
\Psi=B_1b_1+B_2b_2+B_3b_3+Ii+Jj
\end{eqnarray}
into the formula for the potential, we get\footnote{Note that only the composition $b_1\star i\star j$ and  $i\star j\star b_1$ contribute to the potential since the composition is zero or an element of $\mbox{Hom}^*(D4,D4)$ otherwise.}
\begin{eqnarray}
W=\mbox{Tr}\ B_1[B_2,B_3]+JB_1I.
\end{eqnarray}

If we were to introduce a D4-brane of one of the other two orientations, we would end up with the very same quiver but potential  involving terms $JB_2I$ and $JB_3I$ respectively. Looking at a system of intersecting stacks of D4-branes along $\mathbb{C}_{\epsilon_2}\times \mathbb{C}_{\epsilon_3}$ and $\mathbb{C}_{\epsilon_1}\times \mathbb{C}_{\epsilon_3}$, we would get a quiver containing two framing nodes
\[
\begin{tikzcd}
\boxed{k_1} \arrow[rd, bend left, "I^1"]   & & \arrow[ld, bend left,  "I^2"] \boxed{k_2}  \\
 & \arrow[lu, bend left, "J^1"] \arrow[ru, bend left, "J^2"]  \mathcircled{n}\arrow[out=310,in=350,loop,swap,"B_3"]
  \arrow[out=190,in=230,loop,swap,"B_2"]
  \arrow[out=250,in=290,loop,swap,"\normalsize{B_1}"]&
\end{tikzcd}
\]
with $k_i$ denoting multiplicities of D4-branes in each stack. The potential would then contain terms $J^1B_1I^1+J^2B_2I^2$. The most general configuration containing all three stacks of D4-branes would then lead to a quiver with three framing nodes and all the three terms present in the potential. We recover the moduli space of spiked instantons from \cite{Nekrasov:2016qym,Rapcak:2018nsl}.

One might also want to incorporate the dynamics of D4-D4 strings into the story by including an extra generator of $\mbox{Hom}^1(\mbox{D}4[1],\mbox{D}4[1])$ of the form:
\begin{eqnarray}
a\quad :\quad 
\xymatrixcolsep{3pc}
\xymatrixrowsep{3pc}
\xymatrix{ &  & \mathcal{O}\ar[d]^{\scriptsize{1}} \ar[r] &  \mathcal{O} \\
        & \mathcal{O} \ar[r] &  \mathcal{O} & }
\end{eqnarray}
leading to the quiver of the form 
\[
\begin{tikzcd}
\boxed{k}  \arrow[out=160,in=200,loop,swap, "\bullet" marking, "A"] \arrow[r,   shift left=0.5ex, "I"] & \arrow[l, shift left=0.5ex,   "J"]  \mathcircled{n}\arrow[out=340,in=20,loop,swap,"B_3"]
  \arrow[out=70,in=110,loop,swap,"B_2"]
  \arrow[out=250,in=290,loop,swap,"\normalsize{B_1}"]
\end{tikzcd}
\]
and a modified potential 
\begin{eqnarray}
W=\mbox{Tr}\ B_1[B_2,B_3]+JB_1I-IAJ.
\end{eqnarray}
Remember that we should be thinking of $A$ as a non-dynamical field with a prescribed form. To stress the difference compared to the dynamical fields, we label such frozen arrows by a dot in the quiver diagram. We recover the quiver with potential from \cite{Chuang:2019qdz}.

More generally, the modification of the potential with all three framings non-trivial and all the D4-D4 strings incorporated would be
\begin{eqnarray}
\mbox{Tr}\ B_1[B_2,B_3]+\sum_{a}^{3}J^{a}B_aI^{a}-\sum_{a,b=1}^{3}I^aA^{ab}J^{b},
\end{eqnarray}
where we have introduced notation $A^{ab}$ for the maps $A^{ab}\in \mbox{Hom}(\mathbb{C}^{k_a}, \mathbb{C}^{k_b})$ between framing nodes. For example, for two intersecting stacks of D4-branes, the quiver would be of the form
\[
\begin{tikzcd}
\boxed{k} \arrow[out=115,in=155,loop, "\bullet" marking, swap,"A^1"]  \arrow[rd,  "I^1"]  \arrow[rr,  "\bullet" marking, bend left, "A^{12}"]   & & \arrow[ll, "\bullet" marking,  "A^{21}"]  \arrow[ld, bend left,  "I_2"] \arrow[out=25,in=65,loop, "\bullet" marking, swap,"A^2"]   \boxed{k}  \\
 & \arrow[lu, bend left, "J^{23}"] \arrow[ru, "J^{13}"]  \mathcircled{n}\arrow[out=310,in=350,loop,swap,"B_3"]
  \arrow[out=190,in=230,loop,swap,"B_2"]
  \arrow[out=250,in=290,loop,swap,"\normalsize{B_1}"]&
\end{tikzcd}
\]
Such quivers will be explored further in \cite{BR}.

\subsubsection{D0-D2 system}

Let us finish with the discussion of a much-less-explored D0-D2 system. The generators $b_1,b_2,b_3$ are again present as in the previous examples but now we have the following representatives of $\mbox{Hom}^1(\mbox{D}2[1],\mbox{D}0)$ and $\mbox{Hom}^1(\mbox{D}0,\mbox{D}2[1])$ respectively:
\begin{eqnarray}\nonumber
i\quad &:&\quad 
\xymatrixcolsep{3pc}
\xymatrixrowsep{3pc}
\xymatrix{ & \mathcal{O}\ar[d]^{\scriptsize{\begin{pmatrix}0\\0\\1\end{pmatrix}}} \ar[r] & \mathcal{O}^2\ar[d]^{\scriptsize{\begin{pmatrix}-1&0 \\0&-1\\0&0\end{pmatrix}}} \ar[r] &  \mathcal{O}\ar[d]^{\scriptsize{1}} \\
         \mathcal{O} \ar[r] & \mathcal{O}^3 \ar[r] &  \mathcal{O}^3 \ar[r] & \mathcal{O}}\\ \nonumber
&&\vspace{2cm}\\  
j_1\quad &:&\quad 
\xymatrixcolsep{3pc}
\xymatrixrowsep{3pc}
\xymatrix{ &\mathcal{O}\ar[d]^{\scriptsize{\begin{pmatrix}1\\ 0 \end{pmatrix}}} \ar[r] & \mathcal{O}^3\ar[d]^{\scriptsize{\begin{pmatrix}1&0&0\end{pmatrix}}}\ar[r] &  \mathcal{O}^3 \ar[r] & \mathcal{O}  \\
         \mathcal{O} \ar[r] & \mathcal{O}^2 \ar[r] &  \mathcal{O}  & &}\\ \nonumber
&&\vspace{2cm}\\  \nonumber
j_2\quad &:&\quad 
\xymatrixcolsep{3pc}
\xymatrixrowsep{3pc}
\xymatrix{ &\mathcal{O}\ar[d]^{\scriptsize{\begin{pmatrix}0\\ -1 \end{pmatrix}}} \ar[r] & \mathcal{O}^3\ar[d]^{\scriptsize{\begin{pmatrix}0&1&0\end{pmatrix}}}\ar[r] &  \mathcal{O}^3 \ar[r] & \mathcal{O}  \\
         \mathcal{O} \ar[r] & \mathcal{O}^2 \ar[r] &  \mathcal{O}  & &}
\end{eqnarray}
leading to the quiver
\[
\begin{tikzcd}
\boxed{k} \arrow[r, bend left, shift left=0.4ex, "I"] & \arrow[l,   "J_1"]   \arrow[l, shift left=0.4ex, bend left,   "J_2"]   \mathcircled{n}\arrow[out=340,in=20,loop,swap,"B_3"]
  \arrow[out=70,in=110,loop,swap,"B_2"]
  \arrow[out=250,in=290,loop,swap,"\normalsize{B_1}"]
\end{tikzcd}
\]
Inserting 
\begin{eqnarray}
\Psi=B_1b_1+B_2b_2+B_3b_3+Ii+J_1j_1+J_2j_2
\end{eqnarray}
into the formula for the potential, we get
\begin{eqnarray}
W=\mbox{Tr}\ B_1[B_2,B_3]+J_2B_1I-J_1B_2I.
\end{eqnarray}
This example is going to serve as a toy model for constructing BPS algebras associated with framed quivers in later sections. 

The exploration of BPS algebras associated with more complicated D0-D2 systems is a part of an on-going work \cite{BR,RSYZ}. Combining non-compactly supported branes of different dimensions leads to even richer class of examples and will be discussed in \cite{BR}.

\section{Supersymmetric vacua}

\subsection{Quiver quantum mechanics}

In the previous section, we have formulated a model for branes in toric Calabi-Yau three-folds in the language of derived categories of coherent sheaves. We have identified massless string modes between branes $A$ and $B$ with generators of $\mbox{Hom}^*(A,B)$. We have then encoded the spectrum of ghost-one massless string modes in terms of arrows in a quiver diagram and determined the potential capturing the $A_\infty$-structure of morphisms in our brane category. 

We have also stated that the above data uniquely specifies an $\mathcal{N}=4$ gauged quiver quantum mechanics describing the low energy behavior of D0-branes bound to a given background of non-compactly supported branes. The gauge group of such a quantum mechanics consisted of a $U(n_i)$ factor for each of the $m$ circular nodes in the quiver diagram, i.e.
\begin{eqnarray}
G=\bigtimes_{i} U(n_i),
\end{eqnarray}
where the integers $n_i$ indicates the number of branes of a given type and the product runs over all circular (gauge) nodes of the quiver. In the case of $\mathbb{C}^3$ quivers, there is only single circular node with $n$ counting the number of D0-branes and the resulting gauge group is simply $U(n)$. Matter fields $M$ arising from the compactification on  toric Calabi-Yau threefolds consisted of factors 
\begin{eqnarray}
M=\bigoplus_{i} M_i
\end{eqnarray}
for each arrow in the diagram (including those connected to the flavor nodes of the quiver). Each of the factors $M_i$ were identified with $\mbox{Hom}(\mathbb{C}^{n_{t(i)}} \mathbb{C}^{n_{h(i)}})$
where $t(i)$ is the tail node of the arrow $i$ and $h(i)$ is its head node. The gauge group $G$ acts on $M_i$ by
\begin{eqnarray}
g: m_i\rightarrow g_{h(i)}m_ig^{-1}_{t(i)}
\end{eqnarray}
where $g=g_1\times \dots \times g_m$ and $m_i\in M_i$.

In this section, we would like to study the moduli space of vacua of such a quiver quantum mechanics. Roughly speaking, the moduli space of vacua can be identified with the de Rham cohomology of the critical locus of $W$ in the quotient space $M/G$, i.e. the locus minimizing the potential up to gauge transformations. The main problem is non-compactness of such a space. The problem with compactness can be traced back to the D0-branes being able to move escape to infinity. To solve this issue, we introduce the $\Omega$-background (equivariance) that confines D0-branes to the origin. At the level of quiver quantum mechanics, this deformation introduces twisted masses to flavor symmetries of the system that lifts the energy of a large portion of the moduli space of vacua. The de Rham cohomology becomes the equivariant cohomology with the mass parameters identified with equivariant parameters.

\subsection{Twisted masses and equivariance}
\label{sec:flavor}

We are now going to introduce so-called twisted masses \cite{Dorey:1998yh,Ohta:2014ria} for $U(1)$ flavor symmetries of our quantum mechanics. A flavor symmetry is a group acting on $M$ that preserves the potential but that is distinct from the gauge group $G$. For each $U(1)$ factor in our flavor symmetry, we can turn on a mass deformation parametrized by parameter $\mu_i$. Such a deformation arises from gauging the flavor symmetry (introducing a $U(1)$ vector multiplet associated to it) and giving a non-trivial vacuum expectation value $\mu_i$ to scalars of the corresponding vector multiplet.

For example, the vector space $\mathbb{C}^{k_\alpha}$ associated with each framing node of rank $k_\alpha$ can be rotated by $U(k_{\alpha})$ leading to a possible deformation by $k_{\alpha}$ parameters associated with its Cartan subgroup $U(1)^{k_\alpha}\subset U(k_\alpha)$. The potential is generally invariant\footnote{One needs to be careful when introducing a non-trivial nilpotent vacuum expectation value for the Higgs field since such a modification breaks the flavor symmetry to a subtorus \cite{Chuang:2019qdz}. } under such a deformation.

Matter fields $M$ arising from toric Calabi-Yau three-folds admit yet another $U(1)^2$ flavor symmetry. Let us investigate this symmetry in the simplest example of $\mathbb{C}^3$. Note that the transformation  
\begin{eqnarray}
(e^{i\epsilon_1},e^{i\epsilon_2},e^{i\epsilon_3}):B_i\rightarrow e^{i\epsilon_i}B_i
\label{eqib}
\end{eqnarray}
preserves the potential $\mbox{Tr}\ B_1[B_2,B_3]$ if we restrict to the subtorus $U(1)^2\subset U(1)^3$ given by $\epsilon_1+\epsilon_2+\epsilon_3=0$. This flavor symmetry persists also if we introduce framing nodes. For example, the transformation
\begin{eqnarray}\nonumber
(e^{i\epsilon_1},e^{i\epsilon_2},e^{i\epsilon_3}):I&\rightarrow& I,\\
(e^{i\epsilon_1},e^{i\epsilon_2},e^{i\epsilon_3}):J&\rightarrow& e^{i(\epsilon_1+\epsilon_2)}J
\end{eqnarray}
preserves the term $JB_3I$ in the potential associated with the D4-brane along $x_3=0$ and analogously for other orientations and dimensions of the brane. Similar flavor symmetry exists also in quivers arising from other toric Calabi-Yau threefolds. 

Let us trace back the geometric origin of this $U(1)^2$ action. First, $\mathbb{C}^3$ admits an obvious $U(1)^3$ action rotating the three coordinate lines of $\mathbb{C}^3$. Let us introduce the notation $\mathbb{C}_{n_1\epsilon_1+n_2\epsilon_2+n_3\epsilon_3}$ for $\mathbb{C}^3$ together with the information about the $U(1)^3$ action 
\begin{eqnarray}
(e^{i\epsilon_1},e^{i\epsilon_2},e^{i\epsilon_3}):z\rightarrow e^{i(n_1\epsilon_1+n_2\epsilon_2+n_3\epsilon_3)}z
\end{eqnarray}
for $z\in \mathbb{C}_{n_1\epsilon_1+n_2\epsilon_2+n_3\epsilon_3}$. Keeping track of this torus action, our Calabi-Yau threefold is
\begin{eqnarray}
\mathbb{C}_{\epsilon_1}\times \mathbb{C}_{\epsilon_2}\times \mathbb{C}_{\epsilon_3}.
\end{eqnarray}
We can now lift the projective resolutions of all our branes into equivariant complexes. For example, the D0-brane complex lifts into
\begin{eqnarray}\nonumber
\xymatrixcolsep{1.8pc}
\xymatrixrowsep{3pc}
\mathcal{O}_{0}\xrightarrow{\scriptsize{\begin{pmatrix}-x_1\\ x_2\\ -x_3 \end{pmatrix}}}\mathcal{O}_{\epsilon_1}\oplus \mathcal{O}_{\epsilon_2}\oplus \mathcal{O}_{\epsilon_3}\xrightarrow{\scriptsize{\begin{pmatrix}0&-x_3&-x_2\\ -x_3&0&x_1\\ x_2& x_1& 0 \end{pmatrix}}}\mathcal{O}_{\epsilon_2+\epsilon_3}\oplus \mathcal{O}_{\epsilon_1+\epsilon_3}\oplus \mathcal{O}_{\epsilon_1+\epsilon_2}\xrightarrow{\scriptsize{\begin{pmatrix}x_1&x_2&x_3 \end{pmatrix}}}\mathcal{O}_{\epsilon_1+\epsilon_2+\epsilon_3}\\
\end{eqnarray}
where we introduced the notation $\mathcal{O}_{n_1\epsilon_1+n_2\epsilon_2+n_3\epsilon_3}=\mathcal{O}\otimes \mathbb{C}_{n_1\epsilon_1+n_2\epsilon_2+n_3\epsilon_3} $. After the lift, we can see that $B_i$ transforms as $\mathbb{C}_{\epsilon_i}$ since $B_1$ is given by 
\begin{eqnarray}
\xymatrixcolsep{1.7pc}
\xymatrixrowsep{3pc}
\xymatrix{ & \mathcal{O}_0\ar[d]^{\scriptsize{\begin{pmatrix}-1\\0\\0\end{pmatrix}}} \ar[r] & \mathcal{O}_{\epsilon_1}\oplus \mathcal{O}_{\epsilon_2}\oplus \mathcal{O}_{\epsilon_3}\ar[d]^{\scriptsize{\begin{pmatrix}0&0 & 0\\0&0&1\\0&1&0\end{pmatrix}}} \ar[r] &  \mathcal{O}_{\epsilon_2+\epsilon_3}\oplus \mathcal{O}_{\epsilon_1+\epsilon_3}\oplus \mathcal{O}_{\epsilon_1+\epsilon_2}\ar[d]^{\scriptsize{\begin{pmatrix}1,0,0\end{pmatrix}}} \ar[r] & \mathcal{O}  \\
         \mathcal{O} \ar[r] & \mathcal{O}_{\epsilon_1}\oplus \mathcal{O}_{\epsilon_2}\oplus \mathcal{O}_{\epsilon_3} \ar[r] &  \mathcal{O}_{\epsilon_2+\epsilon_3}\oplus \mathcal{O}_{\epsilon_1+\epsilon_3}\oplus \mathcal{O}_{\epsilon_1+\epsilon_2} \ar[r] & \mathcal{O}&}
\vspace{2cm}
\end{eqnarray} 
and it increases the $U(1)^3$ weight by $(1,0,0)$ and analogously for $B_2,B_3$. The origin of the transformation properties (\ref{eqib}) can be thus traced back exactly to the $U(1)^3$ action on $\mathbb{C}^3$.  Similarly, we can identify the transformation properties of $I$ and $J$ maps.

Remember that only the subtorus $U(1)^2$ given by $\epsilon_1+\epsilon_2+\epsilon_3$ was the actual flavor symmetry preserving the potential. This constraint can be though of as a condition on the invariance of the trace map $\int$. Remember that $\int$ simply identified the coefficient in front of a fixed generator of $\mbox{Hom}^3(\mbox{D}0,\mbox{D}0)$. Lifting such a generator to an equivariant map, we find that it transforms as $\mathbb{C}_{\epsilon_1+\epsilon_2+\epsilon_3}$ under the $U(1)^3$ action and we obviously need $\epsilon_1+\epsilon_2+\epsilon_3=0$ for $\int$ to be invariant. 

Let me mention one more perspective on the $\epsilon_1+\epsilon_2+\epsilon_3=0$  constraint. If we were to study a D6-brane on a toric Calabi-Yau threefold, we would identify 
\begin{eqnarray}
\mbox{Hom}^*(\mbox{D}6,\mbox{D}6)=H^{*,0}_{\bar{\partial}}(\mathcal{O}_X)
\end{eqnarray}
with the trace map being the 6d holomorphic Chern-Simons functional \cite{Witten:1992fb}
\begin{eqnarray}
\int \alpha = \int_X \alpha\wedge \Omega
\end{eqnarray}
where $\Omega$ is the Calabi-Yau volume form. In our case 
\begin{eqnarray}
\Omega=dx_1\wedge dx_2\wedge dx_3
\end{eqnarray}
and we see that its invariance requires $\epsilon_1+\epsilon_2+\epsilon_3=0$. This constraint is thus a condition of preserving the Calabi-Yau volume form.

\subsection{Critical equivariant cohomology}

After deriving the quiver with potential to which we associated a supersymmetric quantum mechanics capturing the low energy dynamics of a system of D0-branes bound to higher dimensional non-compact branes, we would like to study their spectrum of BPS vacua. Instead of going through the precise derivation\footnote{Interested reader is encouraged to follow the analysis of \cite{Ohta:2014ria}.} that would require reviewing the explicit construction of the supersymmetric quantum mechanics, studying its topological twists and identifying the cohomology theory describing the BPS states, we are going to simply state the result and justify various terms appearing in the differential.

Since the ground-breaking work of \cite{Witten:1982im} (forgetting about the vector multiplet for a moment), BPS vacua of supersymmetric quantum mechanics can be identified with the de Rham cohomology of the moduli of vacua parametrized by the vacuum expectation value of scalars of chiral multiplets. In our story, these are exactly representations $M$ of the relevant quivers. Furthermore, if the potential $W$ is non-trivial, the differential receives a correction proportional to $dW \wedge$. 

Introducing gauge fields, one needs to "divide by the gauge group". The correct way\footnote{One can equivalently impose vanishing of the moment map of the $G$ action and divide by $G$ itself. See e.g. \cite{mumford1994geometric}.} to divide by the gauge group is to perform so-called geometric-invariant-theory (GIT) quotient.  In the GIT quotient, we need to quotient by the complexified gauge group $G^{\mathbb{C}}$ (factors of $GL(n_i)$ in the examples arising from toric Calabi-Yau threefolds) and impose further stability conditions that makes the quotient well-defined. The stability condition in our $\mathbb{C}^3$ examples simply requires the whole space $\mathbb{C}^{n}$ associated with the circular node to be generated by the action of $B_i$ on $I$'s, i.e. schematically
\begin{eqnarray}
\mathbb{C}^n=\sum_\alpha \mathbb{C}[B_1,B_2,B_3]I^\alpha \mathbb{C}^{k_{\alpha}}.
\label{stability}
\end{eqnarray}
One can also interpret this condition as a requirement that all D0-branes are bound to the non-compact branes. In the situation where the quotient space is not a nice non-singular variety, we can still compute the equivariant cohomology with respect to the complexified gauge-group action. At the level of the BRST operator, this can be implemented by adding an extra term $Q_{gauge}$.

Finally, we can associate a one-parametric deformation of the theory with any $U(1)$ flavor-symmetry factor. As shown e.g. in \cite{Ohta:2014ria}, this modifies the BRST operator by a factor of
\begin{eqnarray}
\mu_i\iota_{X_i},
\end{eqnarray}
where $\mu_i$ is the corresponding twisted-mass parameter and $\iota_{X_i}$ is the outer product with the vector field $X_i$ generating the flavor symmetry. This factor is exactly the factor realizing the de Rham model of the equivariant cohomology as briefly reviewed in the next subsection \cite{Rapcak:2020ueh}.

The above discussion thus suggests that the desired differential has the following form
\begin{eqnarray}
Q=Q_{gauge}+d+ dW\wedge +\sum_{i}\mu_i\iota_{X_i},
\label{dif}
\end{eqnarray}
where $d$ is the standard de Rham differential acting on the chiral-multiplet moduli space (with the stability conditions imposed) and $i$ runs over all the $U(1)$ factors of the flavor symmetry.  The cohomology theory with the differential (\ref{dif}) is known as the de Rham model of the equivariant critical cohomology \cite{Rapcak:2020ueh} and will be denoted as
\begin{eqnarray}
H^*_{U(n)\times U(1)^{k+2}}(M/GL(n),\mbox{Crit}\ W ),
\end{eqnarray}
where $GL(n)$ is the complexified gauge of our quiver quantum mechanics, the factor $k$ in $U(1)^{k+2}$ accounts for the Cartan generators of the flavor symmetry acting on framing nodes and the $+2$ factor to the $\Omega$-background of $\mathbb{C}_{\epsilon_1}\times \mathbb{C}_{\epsilon_2}\times\mathbb{C}_{\epsilon_3}$. Remember also that we restrict to matter fields $M$ satisfying the stability condition (\ref{stability}).

\subsection{Example of equivariant cohomology}

Let us now pause for a little bit and give an example of a calculation of the equivariant cohomology. The reader interested in a detailed introduction of the equivariant cohomology is recommended to consult for example \cite{Guillemin1999SupersymmetryAE,Tu2020}. 

We are going to compute the equivariant cohomology of $\mathbb{C}_{\epsilon}$ with the $U(1)$ action given by $e^{i\epsilon}z$ with $(z,\bar{z})\in \mathbb{C}$ the complex coordinates. The differential is thus of the form
\begin{eqnarray}
Q=dz\partial+d\bar{z}\bar{\partial}+\epsilon \iota_{z\frac{\partial}{\partial z}-\bar{z}\frac{\partial}{\partial \bar{z}}}.
\end{eqnarray}
Since the multiplication by $dz$ and $d\bar{z}$ increases the degree of the differential form by one and $\iota_{z\frac{\partial}{\partial z}-\bar{z}\frac{\partial}{\partial \bar{z}}}$ decreases it by one, if we assign degree two to the parameter $\epsilon$, the full differential $Q$ is of degree one. Unfortunately, when acting on a general form, the differential $Q$ does not square to zero. For example,
\begin{eqnarray}
Q^2z= Q dz= \epsilon z.
\end{eqnarray}
But restricting to $U(1)$-invariant forms, $Q$ is nilpotent and its cohomology makes sense. Generally, when interested in a $G$-equivariant cohomology, one needs to restrict to $G$-invariant forms.  

At degree zero, we have 
\begin{eqnarray}
Qf\left (|z|^2\right ) = \left (\bar{z}dz+zd\bar{z}\right )\frac{\partial f\left (|z|^2\right )}{\partial |z|^2},
\end{eqnarray}
requiring $f$ to be a constant and the cohomology generated by $1$. 

A  general $U(1)$-invariant form at degree one is of the form
\begin{eqnarray}
f\left (|z|^2\right ) zd\bar{z}+g\left (|z|^2\right ) \bar{z}dz.
\end{eqnarray}
The kernel condition requires
\begin{eqnarray}
\left (\left (\frac{\partial f\left (|z|^2\right )}{\partial |z|^2}-\frac{\partial g\left (|z|^2\right )}{\partial |z|^2}\right ) |z|^2+f\left (|z|^2\right )-g\left (|z|^2\right )\right )dz d\bar{z}+\epsilon \left (f\left (|z|^2\right )-g\left (|z|^2\right ) \right )|z|^2
\end{eqnarray}
to vanish that implies $f=g$ but all such elements can be generated by the action of $Q$ on degree-zero forms. 

On the other hand, at degree two, we can write
\begin{eqnarray}
f\left (|z|^2\right )dzd\bar{z}+\epsilon g\left (|z|^2\right )
\end{eqnarray}
and the action of $Q$ produces
\begin{eqnarray}
\epsilon \left ( f\left (|z|^2\right )+\frac{\partial g\left (|z|^2\right )}{\partial }\right ) \left (zd\bar{z}+\bar{z}dz\right ),
\end{eqnarray}
requiring 
\begin{eqnarray}
f(|z|^2)=\frac{\partial g(|z|^2)}{\partial |z|^2}.
\end{eqnarray}
The kernel of $Q$ is thus generated by
\begin{eqnarray}
n|z|^{2(n-1)}dzd\bar{z}+\epsilon |z|^{2n}
\end{eqnarray}
for any $n\geq 0$. On the other hand, the image of $Q$ is generated
\begin{eqnarray}
n|z|^{2n}dz d\bar{z}+\epsilon |z|^{2(n+1)}
\end{eqnarray} 
for any $n\geq 0$ and we can see that the cohomology is generated simply by $\epsilon$. One can proceed analogously at higher degrees and identify 
\begin{eqnarray}
H^*_{U(1)}(\mathbb{C}_{\epsilon})=\mathbb{C}[\epsilon],
\end{eqnarray}
i.e. we get one factor of $\mathbb{C}[\epsilon]$ that is identical to the number  of fixed points of the $U(1)$ action that is also one.

This observation generalizes for an arbitrary equivariant cohomology of a torus action $U(1)^n$ on a manifold with isolated fixed points. According to the Borel localization theorem, different contributions to the equivariant cohomology come from different fixed points $p\in F$ of the torus action, i.e. we have an isomorphism\footnote{The precise claim relates the equivariant cohomology of the space with the equivariant cohomology of the fixed-point set but only up to localization (inversion) of the equivariant parameters.}
\begin{eqnarray}
H^*_{U(1)^n}(X)\rightarrow \oplus_{p\in F} H^*_{U(1)^2}(p) \simeq \oplus_{p\in F} \mathbb{C}[\epsilon_1,\dots, \epsilon_n]|p\rangle.
\end{eqnarray}

This statement receives two corrections in our present situation \cite{kapr,Rapcak:2020ueh}. First, only the fixed points intersecting the critical locus of the potential contribute to the cohomology. Secondly, the fixed-point condition must hold but only up to the gauge transformation.\footnote{I.e. we look for fixed points on the quotient space $M/GL(n)$.} The space of vacua we want to identify is thus in correspondence with fixed-points (up to gauge transformations) of our flavor symmetries lying in the critical locus of the potential. In the rest of this section, we are going to identify such fixed points for quivers associated with a single D2-, D4- and D6-brane respectively. For the analysis in more general situations see \cite{BR}.

\subsection{D2 and 1d partitions}

Let us start with the simplest example of a single framing by one D2-brane. Starting with the potential 
\begin{eqnarray}
W=\mbox{Tr}\ B_1[B_2,B_3]+(J_2B_1-J_1B_2)I ,
\end{eqnarray}
we have the following equations of motion
\begin{eqnarray}\nonumber
{[}B_1,B_3{]}=IJ_1,\qquad {[}B_2,B_3{]}=IJ_2,\qquad {[}B_1,B_2{]}=0\\
B_1I=0,\qquad B_2I=0,\qquad J_2B_1-J_1B_2=0,
\label{eqn2}
\end{eqnarray}
carving out the critical locus.

We can show that in the case of a single D2-brane, these relations imply $J_1=J_2=0$. Since the stability condition requires $\mathbb{C}[B_1,B_2,B_3]I$ to generate the whole $\mathbb{C}^n$, this is equivalent to the vanishing of 
\begin{eqnarray}
J_1P(B_1,B_2,B_3)I=J_2P(B_1,B_2,B_3)I=0
\end{eqnarray}
for any polynomial $P(B_1,B_2,B_3)$.  We are going to prove this by induction in the degree of the polynomial $P(B_1,B_2,B_3)$. First, we obviously have
\begin{eqnarray}
J_iI=Tr IJ_i=\mbox{Tr}\ (B_iB_3-B_3B_i)=0.
\end{eqnarray}
Let us now assume that the above condition holds for all the polynomials of degree $n$. Any $J_iP(B_1,B_2,B_3)I$  with $P(B_1,B_2,B_3)$ being a polynomial of degree $n+1$ can be then written as a sum of monomials of the form $J_iB_1^kB^l_2B^m_3I$ with $k+l+m=n+1$. This is a result of $[B_1,B_2]=0$ and the commutators $[B_1,B_3]$ and $[B_2,B_3]$ leading to products  of lower-degree terms that vanish. In particular
\begin{eqnarray}
J_i P_1(B_1,B_2,B_3)[B_j,B_3]P_2(B_1,B_2,B_3)I&=&J_i P_1(B_1,B_2,B_3)IJ_jP_2(B_1,B_2,B_3)I
\end{eqnarray} 
that obviously vanishes due to the induction hypothesis. We then have
\begin{eqnarray}\nonumber
J_1B_1^kB_2^lB_3^mI&=&Tr B_1^kB_2^lB_3^m(B_1B_3-B_3B_1)=-Tr B_1^kB_2^l[B_1,B_3^m]B_3\\
&=&-\sum_{r=0}^{m-1}Tr B_1^kB_2^lB_3^{r} IJ_1B_3^{m-r}=-\sum_{n=0}^{m-1}J_1 B_3^{m-r}B_1^kB_2^lB_3^{r} I\\ \nonumber
&=&-mJ_1B_1^kB_2^lB_3^mI,
\end{eqnarray}
where we use the re-ordering argument above and the combination $J_1B_1^kB_2^lB_3^mI$ must vanish as well. Using an analogous argument for $J_2$, we see that we can always set $J_1=J_2=0$.

The system thus reduces to the system of mutually commuting $B_1,B_2,B_3$ and an extra $I$. We can furthermore set $B_1=B_2=0$ since
\begin{eqnarray}
B_1\mathbb{C}^n=B_1\mathbb{C}[B_1,B_2,B_3]I=\mathbb{C}[B_1,B_2,B_3]B_1I=0
\end{eqnarray}
and analogously for $B_2$ as a consequence of the commutativity of $B_1,B_2,B_3$, the stability condition and conditions $B_1I=B_2I=0$. We thus reduced our problem to the system to a single matrix $B_3$ with stability condition
\begin{eqnarray}
\mathbb{C}^n=\mathbb{C}[B_1]I.
\end{eqnarray}

The moduli of vacua (before turning on twisted masses) is thus parametrized by a pair $(B_1,I)$ satisfying the stability condition above modulo gauge transformations
\begin{eqnarray}
g: (B_1,I)\rightarrow (gB_1g^{-1},gI)
\end{eqnarray}
for any $g\in GL(n)$. We need to now identify fixed points of the $U(1)^2$ flavor group acting on the involved matrices as
\begin{eqnarray}
(e^{i\epsilon_1},e^{i\epsilon_2},e^{i\epsilon_3}): (B_1,I)\rightarrow (e^{i\epsilon_1}B_1,I).
\end{eqnarray}

To gain some experience with finding fixed points, let us start with a very explicit analysis for $n=1$. First, the value of $I$ is non-vanishing due to the stability condition. It can be thus set to $1$ by the gauge transformation. The fixed-point condition then requires
\begin{eqnarray}
e^{i\epsilon_1}B_1=gB_1g^{-1}=B_1,
\end{eqnarray} 
leading to $B_1=0$. The only fixed point can be thus identified with the gauge orbit of $(B_1,I)=(0,1)$

Moving to $n=2$, we have $I\in \mbox{Hom}(\mathbb{C},\mathbb{C}^2)$ and  $B_1\in \mbox{Hom}(\mathbb{C}^2,\mathbb{C}^2)$. $I$ being non-zero due to the stability condition, the gauge transformation allows us to fix
\begin{eqnarray}
I=\begin{pmatrix}
1\\ 0
\end{pmatrix}.
\end{eqnarray}
After such a fixing, the subgroup of $GL(2)$ parametrized by
\begin{eqnarray}
g=\begin{pmatrix}
1& b \\ 0 & d
\end{pmatrix}
\end{eqnarray}
preserves this choice. Adjusting $b$ in the gauge transformation above, we can furthermore set
\begin{eqnarray}
B_1=\begin{pmatrix}
\alpha & 0 \\ \beta & \gamma
\end{pmatrix}
\end{eqnarray}
with the remaining gauge freedom parametrized by
\begin{eqnarray}
g=\begin{pmatrix}
1& 0 \\ 0 & d
\end{pmatrix}.
\label{restrict}
\end{eqnarray}
Let us now impose the fixed-point condition
\begin{eqnarray}
e^{i\epsilon_1}B_1=e^{\epsilon_1}\begin{pmatrix}
\alpha & 0 \\ \beta & \gamma
\end{pmatrix}= gB_1g^{-1}.
\end{eqnarray} 
The restricted gauge transformation $g$ from (\ref{restrict}) can only rescale the $\beta$ component and we thus need $\alpha=\gamma=0$. On the other hand, $\beta\neq 0$ from the stability condition and we can use the remaining gauge transformation to fix
\begin{eqnarray}
(B_1,I)=\left (\begin{pmatrix}
0& 0\\ 1 & 0
\end{pmatrix},
\begin{pmatrix}
1\\ 0
\end{pmatrix}\right ).
\end{eqnarray}
The gauge orbit of this pair is the only fixed point of the $n=2$ moduli. The value of $d$ in  (\ref{restrict}) is then $e^{i\epsilon_1}$.

Let us now investigate fixed points for general $n$. The condition of $(B_1,I)$ being a fixed point (modulo gauge transformation) requires an existence of $g$ such that
\begin{eqnarray}
e^{i\epsilon_1}B_1=gB_1q^{-1},\qquad I=gI.
\end{eqnarray}
Let us choose a basis of $\mathbb{C}^n$ that diagonalizes\footnote{Note that this was the case also in our $n=2$ example.} $g$. Let $a$ be one of the basis vectors with eigenvalue $e^{i(n_1\epsilon_1+n_2\epsilon_2+n_3\epsilon_3)}$ for some $n_i$. We then have
\begin{eqnarray}
g B_1a=g B_1g^{-1}ga =e^{i((n_1+1)\epsilon_1+n_2\epsilon_2+n_3\epsilon_3)} B_1a
\end{eqnarray}
and $B_1a$ is another eigenvector of $g$ with an eigenvalue $e^{i((n_1+1)\epsilon_1+n_2\epsilon_2+n_3\epsilon_3)}$. Analogously, $B_1^{m}a$ is an eigenvector with eigenvalue $e^{i((n_1+m)\epsilon_1+n_2\epsilon_2+n_3\epsilon_3)}$. Finally, since $I$ does not transfer under the $U(1)^2$ action, $I$ is itself one of the eigenvectors. But because the whole $\mathbb{C}^n$ can be generated from $I$ by the action of $B_1$, we can see that in our basis $I,B_1I,B_1^2I,\dots$, the map $B_1$ has generally the form of a nilponent matrix with entries $1$ below diagonal. For example, for $n=4$,
\begin{eqnarray}
B_1=\begin{pmatrix}
0&0&0&0\\
1&0&0&0\\
0&1&0&0\\
0&0&1&0\\
\end{pmatrix}.
\end{eqnarray}

It will be convenient to draw the decomposition of the vector space $\mathbb{C}^n$ as an equivariant complex complex. E.g. in the $n=4$ example,
\begin{eqnarray}
\xymatrixcolsep{3pc}
\xymatrixrowsep{3pc}
\xymatrix{  \mathbb{C}_{0} \ar[r]^{B_1} & \mathbb{C}_{\epsilon_1} \ar[r]^{B_1} & \mathbb{C}_{2\epsilon_1} \ar[r]^{B_1}  & \mathbb{C}_{3\epsilon_1}}
\end{eqnarray}
with the subscript labeling the the equivariant degree under the $U(1)^2$ action as introduced in section \ref{sec:flavor}. Finally, note that we can visualize the weight decomposition of $\mathbb{C}^n$ as a row of $n$ boxes. For example, in our case of $n=4$, 
\begin{eqnarray}
\yng(4)
\end{eqnarray}
with the box at position $n$ in the horizontal direction associated with the subspace $\mathbb{C}_{n\epsilon_1}$. Examples associated with D2-brane framings will be further explored in \cite{RSYZ,BR}.

\subsection{D4 and 2d partitions}

We can proceed in the very same way in the case of the D4-brane framing. The system of equations following from the variation of the potential is now
\begin{eqnarray}\nonumber
{[}B_1,B_2{]}&=&IJ,\\
{[}B_1,B_3{]}&=&{[}B_2,B_3{]}=0,\\ \nonumber
JB_3&=&B_3I=0.
\label{eqn1}
\end{eqnarray}

Note that we now have
\begin{eqnarray}
B_3\mathbb{C}^n=B_3\mathbb{C}[B_1,B_2,B_3]I=\mathbb{C}[B_1,B_2,B_3]B_3I=0
\end{eqnarray}
as a consequence of $\mathbb{C}[B_1,B_2,B_3]I$ generating the whole $\mathbb{C}^n$, $B_3$ commuting with all $B_i$ and the condition $B_1I=0$. We can thus set $B_3=0$ and arrive at the famous ADHM moduli space. 

We can also show that in the rank-one case, we have $J=0$. Since $\mathbb{C}^n=\mathbb{C}[B_1,B_2]I$ due to the stability condition, this is equivalent to showing that $JP(B_1,B_2)I=0$ for any polynomial $P(B_1,B_2)$. Let us prove it inductively in the degree of $P(B_1,B_2)$. We obviously have
\begin{eqnarray}
JI=\mbox{Tr}\ [B_1,B_2]=0.
\end{eqnarray}
If $JP(B_1,B_2)I$ vanishes for all degree $n-1$ polynomials, vanishing at degree $n$ is equivalent to $JB_1^kB_2^lI=0$ for $k+l=n$ since commutation of $B_1,B_2$ produces
\begin{eqnarray}
JP_1(B_1,B_2)[B_1,B_2]P_2(B_1,B_2)I=JP_1(B_1,B_2)IJP_2(B_1,B_2)I
\end{eqnarray}
that is a product of two lower-degree terms that vanish by induction. Finally, we have
\begin{eqnarray}
JB_1^kB_2^lI&=&Tr B_1^kB_2^l(B_1B_2-B_2B_1)=Tr [B_2,B_1^k]B_2^l\\ \nonumber
&=&-\sum_{m=0}^{k-1}B_1^mIJ B^{k-m-1}B_2=-\sum_{m=0}^{k-1}J B^{k-m-1}B_2B_1^mI=-kJB_1^kB_2^lI
\end{eqnarray}
and $JB_1^kB_2^lI$ must vanish.

Analogously to the discussion above, $(B_1,B_2,I)$ being a fixed point requires an existence of $g$ such that
\begin{eqnarray}\nonumber
e^{i\epsilon_1}B_1=gB_1g^{-1},\qquad e^{i\epsilon_2}B_2=gB_2g^{-1},\qquad I=gI.
\end{eqnarray}
Let us pick a basis of $\mathbb{C}^n$ that diagonalizes $g$. We can again see that if $a$ is an eigenvector of $g$ with eigenvalue $e^{i(n_1\epsilon_1+n_2\epsilon_2+n_3\epsilon_3)}$, then $B_1a$ is an eigenvector with eigenvalue $e^{i((n_1+1)\epsilon_1+n_2\epsilon_2+n_3\epsilon_3)}$ and $B_2a$ is an eigenvector with eigenvalue $e^{i(n_1\epsilon_1+(n_2+1)\epsilon_2+n_3\epsilon_3)}$. Furthermore, since the whole $\mathbb{C}^n$ can be generated by an action of $B_1,B_2$ on $I$ and these two mutually commute, we can see that the space $\mathbb{C}^n$ decomposes according to the $U(1)^2$ weights into subspaces specified by a Young diagram.

For example,
\begin{eqnarray}
\yng(1,4)
\end{eqnarray}
would be associated with decomposition
\begin{eqnarray}
\xymatrixcolsep{3pc}
\xymatrixrowsep{3pc}
\xymatrix{  0\ar[r]^{B_1} &  0\ar[r]^{B_1}&  0\ar[r]^{B_1}  &0  \\
      \mathbb{C}_{\epsilon_2}\ar[r]^{B_1} \ar[u]^{B_2}&0\ar[r]^{B_1} \ar[u]^{B_2}&0 \ar[r]^{B_1} \ar[u]^{B_2}&0 \ar[u]^{B_2}  \\
\mathbb{C}_0\ar[r]^{B_1} \ar[u]^{B_2} &  \mathbb{C}_{\epsilon_1} \ar[r]^{B_1} \ar[u]^{B_2}&  \mathbb{C}_{2\epsilon_1}\ar[r]^{B_1} \ar[u]^{B_2} &0  \ar[u]^{B_2}}.
\end{eqnarray}
It is easy to check that this corresponds to the gauge orbit of
\begin{eqnarray}(B_1,B_2,I)=\left (
\begin{pmatrix}
0 & 0 &0 &0 \\
1 & 0 &0 &0 \\
0 & 1 &0 &0 \\
0 & 0 &0 &0 \\
\end{pmatrix},
\begin{pmatrix}
0 & 0 &0 &0 \\
0 & 0 &0 &0 \\
0 & 0 &0 &0 \\
1 & 0 &0 &0 \\
\end{pmatrix},
\begin{pmatrix}
1\\
0\\
0\\
0
\end{pmatrix} \right )
\end{eqnarray}
decomposing $\mathbb{C}^4$  exactly as shown in the diagram. One can proceed analogously with more complicated partitions and see that the fixed points are in correspondence with standard 2d partitions. In particular, a box at position $(n_1,n_2)$ is associated with a factor $\mathbb{C}_{n_1\epsilon_1+n_2\epsilon_2}$.

Note that the decomposition diagram cannot have any holes, i.e. if the subspace of weight $n_1\epsilon_1+n_2\epsilon_2$ is nontrivial, so are subspaces of weight  $(n_1-1)\epsilon_1+n_2\epsilon_2$ and  $n_1\epsilon_1+(n_2-1)\epsilon_2$. If that was not the case and  the commutative diagram would have a hole, there would exists a $g$ eigenvector $a$ such that $B_1a=0$ and $B_1B_2a\neq 0$ (or $B_2a=0$ and $B_2B_1a\neq 0$). But form the commutativity, we would get a contradiction
\begin{eqnarray}
0=B_2B_1a=B_1B_2a\neq 0.
\end{eqnarray}
See \cite{nakajima,article3} for the $\mathbb{C}^2$ perspective and \cite{Nishinaka:2013mba,2019,Rapcak:2018nsl} for the dimensional reduction from $\mathbb{C}^3$.

\subsection{D6 and 3d partitions}

\begin{figure}
  \centering
    \includegraphics[width=0.1\textwidth]{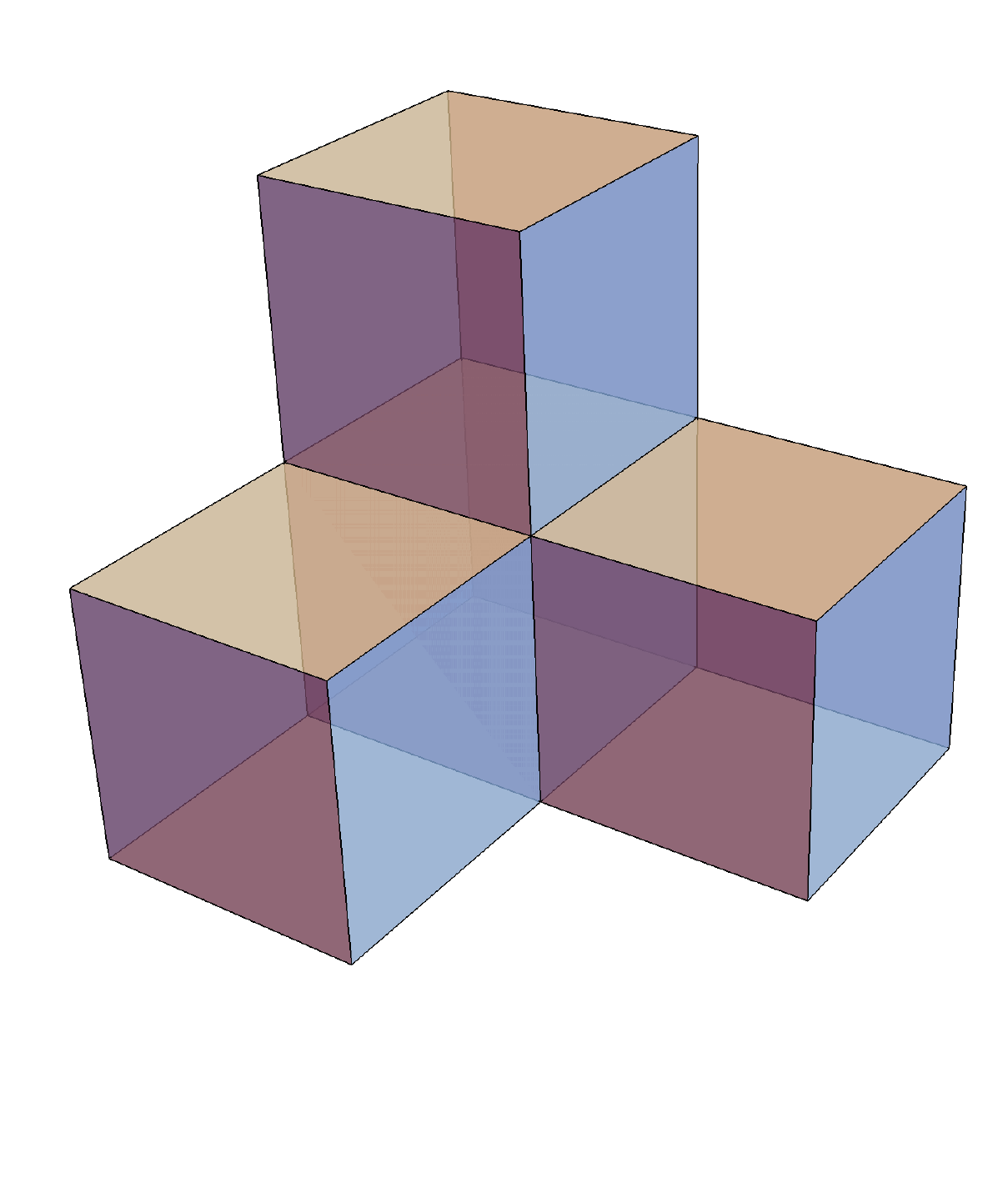}
\caption{An example of a 3d partition indicating the decomposition of $\mathbb{C}^4$ into $\mathbb{C}_{0}\oplus\mathbb{C}_{\epsilon_1}\oplus\mathbb{C}_{\epsilon_2}\oplus\mathbb{C}_{\epsilon_3}$ as a $U(1)^2$-module for the flavor symmetry.}
\label{fig3}
\end{figure}

In the case of the D6-brane framing, we do not have any arrows going to the framing vertex and all the $B_i$ mutually commute. Decomposition of the vector space $\mathbb{C}^n$ into eigenspaces\footnote{Compared to the previous two examples, one needs to be a bit careful when analyzing fixed points of the subtorus $U(1)^2$ defined by $\epsilon_1+\epsilon_2+\epsilon_3=0$. With a little bit of care, one can argue that the fixed points are indeed parametrized by 3d partitions.} of $g$ then leads to the identification of fixed points with 3d partitions. For example, the 3d partition depicted in figure \ref{fig3} would correspond to the gauge orbit of
\begin{eqnarray}
(B_1,B_2,B_3,I)=
\left (
\begin{pmatrix}
0 & 0 &0 &0 \\
1 & 0 &0 &0 \\
0 & 0 &0 &0 \\
0 & 0 &0 &0 \\
\end{pmatrix},
\begin{pmatrix}
0 & 0 &0 &0 \\
0 & 0 &0 &0 \\
1 & 0 &0 &0 \\
0 & 0 &0 &0 \\
\end{pmatrix},
\begin{pmatrix}
0 & 0 &0 &0 \\
0 & 0 &0 &0 \\
0 & 0 &0 &0 \\
1 & 0 &0 &0 \\
\end{pmatrix},
\begin{pmatrix}
1\\
0\\
0\\
0
\end{pmatrix} \right ).
\end{eqnarray}
See \cite{Maulik:2003rzb,Jafferis:2007sg,Cirafici:2008sn} for details.

\section{Modules from correspondences}

\subsection{Correspondences}

After identifying BPS vacua of our supersymmetric quantum mechanics describing the low energy dynamics of D0-branes bound to non-compactly supported branes, we are going to construct a geometric action of two copies of the cohomological Hall algebra (raising and lowering generators) increasing and decreasing the number of D0-branes as depicted in figure \ref{fig2}. In this subsection, we are going to construct the correspondences  for framed moduli spaces associated with a single D2-, D4- and D6-brane respectively. A brief exploration of algebraic structures appearing in other configurations is left for the last chapter.

In particular, we are now going to define an action of two sets of generators 
\begin{eqnarray}\nonumber
e_m:H^*_{U(1)^2}(M(n),\mbox{Crit}(W))\rightarrow H^*_{U(1)^2}(M(n+1),\mbox{Crit}(W)),\\
f_m:H^*_{U(1)^2}(M(n+1),\mbox{Crit}(W))\rightarrow H^*_{U(1)^2}(M(n),\mbox{Crit}(W)),
\end{eqnarray}
where $M(n)=\mathcal{M}(n)/GL(n)$ and $\mathcal{M}(n)$ is given by the quiver representations
\begin{eqnarray}
(B_1,B_2,B_3,I),\qquad (B_1,B_2,B_3,I,J),\qquad (B_1,B_2,B_3,I,J_1,J_2),
\end{eqnarray} 
subject to the stability condition and the circular node of rank $n$. Introducing generators
\begin{eqnarray}
\psi_{m+n}=[e_m,f_n],
\end{eqnarray}
the set $e_n,f_n,\psi_n$ can be shown to satisfy relations of the (shifted) $\mathfrak{gl}_1$ affine Yangian and different choices of framings lead to its different representations \cite{Rapcak:2020ueh,BR}. 

A crucial role in the construction of $e_m,f_m$ is played by a correspondence $M(n+1,n)$ between $M(m+1)$ and $M(n)$, i.e. a closed subset $M(n+1,n)$ in $M(n+1)\times M(n)$ \cite{Kontsevich:2010px,Soibelman:2014gta}. See also \cite{nakajima,Nakajima:1994nid,Nakajima:1995ka,2011dd,schiffmann2012cherednik,2017ee,Rapcak:2018nsl,Rapcak:2020ueh}. Let us start with a definition of such a correspondence for moduli spaces $\mathcal{M}(m+1)$ and $\mathcal{M}(n)$ before quotienting by the gauge group. Let us also restrict to the D4-brane moduli since the only difference between D2-, D4- and D6-brane moduli spaces is a different number of maps $J$. A point 
\begin{eqnarray}
\left (\left (B^{(1)}_1,B_2^{(1)},B_3^{(1)},I^{(1)},J^{(1)}\right ),\left (B^{(2)}_1,B^{(2)}_2,B^{(2)}_3,I^{(2)},J^{(2)}\right )\right )\in \mathcal{M}(n+1)\times\mathcal{M}(n)
\end{eqnarray}
lies in $\mathcal{M}(n,n+1)$ if there exists a map $\xi:\mathbb{C}^{n+1}\rightarrow \mathbb{C}^{n}$ such that
\begin{eqnarray}
\xi B^{(1)}_i=B^{(2)}_i\xi,\qquad \xi I^{(1)}=I^{(2)},\qquad J^{(1)}_i=J^{(2)}_{i}\xi.
\label{constraint}
\end{eqnarray}
The stability condition implies that $\xi$ is a surjective map and $S=\mbox{Ker}\ \xi$ is a one-dimensional subspace of $\mbox{Ker}\  j^{(1)}_i$ that is invariant under the action of $B_i^{(1)}$. We can thus view $\mathcal{M}(n+1,n)$ as a pair of a point in $\mathcal{M}(n+1)$ together with a choice of a subspace $S \subset \mathbb{C}^n$ preserved by the action of $B_i^{(1)}$ and lying in the kernel of $j^{(1)}$. Using this description, we can quotient $\mathcal{M}(n+1,n)$ by the obvious action of $GL(n+1)$ and write 
\begin{eqnarray}
\xymatrix{ & M(n+1,n)\ar[ld]_p \ar[rd]^q &   \\
        M(n+1) & & M(n)   },
\end{eqnarray}
where the map $p$ is the obvious map forgetting the information about the subspace $S$ and $q$ is a quotient of $M(n+1)$ by the subspace $S$. Note also that $S=\mbox{Ker}\ \xi$ gives rise to a line bundle $L$ on the correspondence and will be called the tautological line bundle. 

Starting with $|\alpha \rangle \in H^*_{U(1)^2}(M(n),\mbox{Crit}(W))$, we can now define an operator
\begin{eqnarray}
e_0|\alpha \rangle = p_* \circ q^*|\alpha \rangle 
\end{eqnarray}
by pulling it back by $q$ and pushing forward\footnote{The push-forward in equivariant cohomology can be though of as a fiber integration.} by $p$, obtaining an element in $H^*_{U(1)^2}(M(n+1),\mbox{Crit}(W))$. Reversing the order of the two maps, we have
\begin{eqnarray}\nonumber
f_0|\alpha \rangle &=& q_* \circ p^*|\alpha \rangle 
\end{eqnarray}
for any $|\alpha \rangle \in H^*_{U(1)^2}(M(n+1),\mbox{Crit}(W))$. More generally, utilizing the tautological line bundle, we can define operators 
\begin{eqnarray}\nonumber
e_m|\alpha \rangle &=& p_*   \circ c_1(L)^m\wedge q^*|\alpha \rangle, \\
f_m|\alpha \rangle &=& q_* \circ c_1(L)^m \wedge   p^* |\alpha \rangle 
\end{eqnarray}
for any integer $n$ by multiplying the image of the pull-back by the $n$'th power of the first Chern class of the tautological line bundle $L$ defined above.

Before proceeding with finding explicit formulas for operators $e_m$ and $f_m$, let us pause and discuss fixed points of the correspondence $M(n+1,n)$. As we have seen above, fixed points of $M(n+1)$ are in correspondence with partitions of various dimensions containing $n+1$ boxes. In order to specify a point on $M(n+1,n)$, we need to further identify a subspace of $\mathbb{C}^{n+1}$ that is fixed by the action of $B_i^{(1)}$ and lies in the kernel of $j^{(1)}$. Since $j^{(1)}=0$ at all fixed points of all three of our moduli spaces, we only require  the subspace to be fixed under $B_i^{(1)}$. When restricting to a fixed point $\lambda \in M(n+1)$, we picked a natural basis of $\mathbb{C}^{n+1}$ that diagonalizes $g$ defined by
\begin{eqnarray}
e^{i\epsilon_i}B_i^{(1)} =gB_i^{(1)}g^{-1},\qquad I^{(1)}=gI^{(1)}.
\end{eqnarray}
Basis vectors of this basis were in correspondence with boxes in the partition\footnote{Remember that the partition is 1d, 2d and 3d for D2-, D4- and D6-brane respectively.} $\lambda$ labeling the fixed point. Matrices $B^{(1)}_i$ acting on a vector labeled by a given box $\Box$ produced a vector labeled by a neighboring box in the $i$'th direction. We can thus see that the only one-dimensional subspaces of $\mathbb{C}^{n+1}$ preserved by the action of $B^{(1)}_i$ are those associated with corners of the partition $\lambda$. Fixed points of $M(n+1,n)$ are thus labeled by a pair of partitions with $n+1$ and $n$ boxes, mutually related by an addition/removal of one box. We are going to label such a pair by $(\lambda+\Box,\lambda)$. Note also that the equivariant weight of the vector space associated with the added/removed vector is $\epsilon_{\Box}=n_1\epsilon_1+n_2\epsilon_2+n_3\epsilon_3$, where $(n_1,n_2,n_3)$ are the coordinates of the added/removed box. This factor is going to play an important role in the construction of $e_m$ and $f_m$.

For example, fixed points of $\mathcal{M}(3,2)$ for the D4-brane moduli are given by pairs
\begin{eqnarray}
(\yng(1,1,1),\yng(1,1)),\quad (\yng(1,2),\yng(1,1)),\quad (\yng(1,2),\yng(2)),\quad (\yng(3),\yng(2)).
\end{eqnarray}
The maps $p$ and $q$ project onto the first/second component and give a fixed point of $M(3)$ and $M(2)$ respectively. We label such restricted maps by $p_F$ and $q_F$. Weights of the added/removed box in our examples are $2\epsilon_2, \epsilon_1, \epsilon_2$ and $2\epsilon_1$ respectively.

Remember that we have an isomorphism of equivariant cohomologies roughly of the form\footnote{Remember that the precise claim relates the equivariant cohomology of the space with the equivariant cohomology of the fixed-point set but only up to localization (inversion) of the equivariant parameters. The equivariant cohomology of the fixed-point set then splits into contributions from each point denoted by $\mathbb{C}[\epsilon_1,\epsilon_2,\epsilon_3]|\lambda \rangle$ and $ \mathbb{C}[\epsilon_1,\epsilon_2,\epsilon_3]|\lambda+\Box ,\lambda \rangle $.}
\begin{eqnarray}\nonumber
\oplus_{\lambda \in F_n}\mathbb{C}[\epsilon_1,\epsilon_2,\epsilon_3]|\lambda \rangle &\rightarrow&  H^*_{U(1)^2}(M(n),\mbox{Crit} (W)),\\
\oplus_{(\lambda+\Box,\lambda)\in F_{n+1,n}} \mathbb{C}[\epsilon_1,\epsilon_2,\epsilon_3]|\lambda+\Box ,\lambda \rangle  &\rightarrow&  H^*_{U(1)^2}(M(n+1,n),\mbox{Crit} (W))
\end{eqnarray}
with $F_n$ being the fixed-point set of $M(n)$ and $F_{n+1,n}$ the fixed-point set of $M(n+1,n)$. We have also introduced basis vectors of $H^*_{U(1)^2}(\lambda)$ and $H^*_{U(1)^2}(\lambda+\Box ,\lambda)$ labeled by $|\lambda \rangle$ and $|\lambda+\Box,\lambda \rangle $.  

Obviously, we also have embedding maps of fixed points
\begin{eqnarray}\nonumber
\iota_{\lambda}&:&\lambda \hookrightarrow M(n)\ \mbox{for}\ \lambda \in F_n,\\
\iota_{\lambda,\lambda+\Box}&:&(\lambda+\Box ,\lambda) \hookrightarrow M(n+1,n)\ \mbox{for}\ (\lambda+\Box,\lambda) \in F_{n+1,n}.
\end{eqnarray}
Pushing forward\footnote{The image of the push-forward map can be defined as a top form with a support near the fixed point that integrates to one. This map is known as the equivariant Thom isomorphism \cite{bott2013differential,Costenoble1992TheET}.} the generator $|\lambda \rangle \in H^*_{U(1)^2}(\lambda)$ and $|\lambda+\Box ,\lambda \rangle \in H^*_{U(1)^2}((\lambda+\Box ,\lambda))$ by these maps thus produces a natural (fixed-point) basis of $H^*_{U(1)^2}(M(n),\mbox{Crit} (W))$ and $H^*_{U(1)^2}(M(n+1,n),\mbox{Crit} (W))$ respectively. We also define 
\begin{eqnarray}
\iota^{*}_{F_{n}}=\sum_{\lambda \in F_n}\iota^*_{\lambda},\qquad \iota^{*}_{F_{n+1,n}}=\sum_{(\lambda+\Box,\lambda) \in F_{n+1,n}}\iota^*_{\lambda+\Box,\lambda}.
\end{eqnarray}

We can now consider the following diagram 
\begin{eqnarray}
\xymatrix{ & F_{n+1,n} \ar[ld]_{p_F} \ar[rd]^{q_F} \ar[d]_{\iota_{n+1,n}} &   \\
F_{n+1} \ar[d]_{\iota_{n+1}} & M(n+1,n) \ar[ld]_p \ar[rd]^q &   F_n \ar[d]_{\iota_{n}} \\
        M(n+1) & & M(n)   }.
\end{eqnarray}
Using the above correspondence, we can construct an action operators $e_m,f_m$ in the fixed-point basis 
\begin{eqnarray}
e_m:H^*_{U(1)^2}(F_n)\rightarrow H^*_{U(1)^2}(F_{n+1}),\qquad f_m:H^*_{U(1)^2}(F_{n+1})\rightarrow H^*_{U(1)^2}(F_{n})
\end{eqnarray}
by formulas
\begin{eqnarray}\nonumber
e_m|\lambda \rangle &=& i_{n+1*}^{-1}\circ p_* \circ c_1(L^m) \wedge q^* \circ  i_{n*} |\lambda \rangle,\\
f_m |\lambda \rangle &=&i_{n*}^{-1}\circ q_* \circ c_1(L^m) \wedge p^* \circ  i_{n+1*} |\lambda \rangle.
\end{eqnarray}

To make the formulas more explicit, we need to recall the Atiyah-Bott localization formula in equivariant cohomology \cite{ATIYAH19841}. Let $\lambda$ be a fixed point in $M(n)$. According to the Atiyah-Bott localization formula, we can invert the push-forward of the embedding $\iota_{\lambda}: \lambda \hookrightarrow M(n)$ as
\begin{eqnarray}
\iota_{\lambda*}^{-1} =\frac{\iota^*_{\lambda}}{e_{U(1)^2}(T^*_{\lambda}M(n))},
\end{eqnarray}
where $T^*_{\lambda}M(n)$ is the tangent space of $M(n)$ at a point $\lambda$, $e_{U(1)^2}(T^*_{p}M(n))$ is its Euler character to be described momentarily and $\iota^*_{\lambda}$ is the standard pull-back. Similarly, we can write down the Atiyah-Bott localization formula for the correspondence by exchanging $\lambda \rightarrow \lambda+\Box$ and $M(n)\rightarrow M(n+1,n)$.

The tangent space of $M(n)$ at a fixed point is preserved by the equivariant $U(1)^2$ action and splits into a diret sum of $U(1)^2$ representations
\begin{eqnarray}
T_{\lambda}^*M(n)= \oplus_{\alpha=1}^{\mbox{dim}\ M(n)} \mathbb{C}_{\epsilon_\alpha},
\end{eqnarray}
where $ \mathbb{C}_{\epsilon_\alpha}$ is a complex line transforming as
\begin{eqnarray}
(e^{i\epsilon_1},e^{i\epsilon_2},e^{i\epsilon_3}):z\rightarrow e^{i\epsilon_{\alpha}}z
\end{eqnarray}
for any $z\in \mathbb{C}_{\epsilon_\alpha}$. The Euler character is then given by the product of the equivariant weights
\begin{eqnarray}
e_{U(1)^2}(T_\lambda (\mathcal{M})))=\prod_{\alpha=1}^{\mbox{dim}\ M(n)}\epsilon_{\alpha}.
\end{eqnarray} 

We can now find explicit formulas for the action of $e_m,f_m$ in the fixed point basis. Let us start with the computation for $e_0$. First, we can commute the push-forward maps as
\begin{eqnarray}
e_0|\lambda \rangle &=& p_{F*} \circ  i_{n+1,n*}^{-1} \circ  q^* \circ  i_{n*}|\lambda \rangle.
\end{eqnarray}
The Atiyah-Bott localization formula  then leads to
\begin{eqnarray}
e_0 |\lambda \rangle &=& \sum_{(\mu+\Box,\mu)\in F_{n+1,n}} \frac{1}{e_{U(1)^2}(T^*_{\mu+\Box,\mu}M(n+1,n))} p_{F*} \circ  i_{\mu+\Box,\mu}^{*} \circ  q^* \circ  i_{\lambda *} |\lambda \rangle
\end{eqnarray}
Commuting the pull-back maps gives
\begin{eqnarray}
&=& \sum_{(\mu+\Box,\mu)\in F_{n+1,n}}\frac{1}{e_{U(1)^2}(T^*_{\mu+\Box,\mu}M(n+1,n))} p_{F*} \circ q^*_{\mu+\Box,\mu}  \circ   i_{\mu}^{*}\circ  i_{\lambda*} |\lambda \rangle.
\end{eqnarray}
We can now use the Atiyah-Bott localization formula again to rewrite the pull-back in terms of the inverse map
\begin{eqnarray}
&=& \sum_{(\mu+\Box,\mu)\in F_{n+1,n}} \frac{e_{U(1)^2}(T^*_{\mu}M(n))}{e_{U(1)^2}(T^*_{\mu+\Box,\mu}M(n+1,n))} p_{F*} \circ q^*_{\mu+\Box,\mu}  \circ   i_{\mu*}^{-1}\circ  i_{\lambda*} |\lambda \rangle.
\end{eqnarray}
The composition of $\iota_{\mu*}^{-1}$ with $\iota_{\lambda*}$ vanishes unless $\lambda=\mu$ for which it gives the identity. We can thus simplify as
\begin{eqnarray}
&=& \sum_{(\lambda+\Box,\lambda)\in F_{n+1,n}} \frac{e_{U(1)^2}(T^*_{\lambda}M(n))}{e_{U(1)^2}(T^*_{\lambda+\Box,\lambda}M(n+1,n))} p_{F*} \circ q^*_{\lambda+\Box,\lambda} |\lambda \rangle,
\end{eqnarray}
where the sum goes only over the pairs $(\lambda+\Box,\lambda)$ in $F_{n+1,n}$ for $\lambda$ fixed by the initial state. This leads us to the final expression
\begin{eqnarray}
&=& \sum_{(\lambda+\Box,\lambda)\in F_{n+1,n}} \frac{e_{U(1)^2}(T^*_{\lambda}M(n))}{e_{U(1)^2}(T^*_{\lambda+\Box,\lambda}M(n+1,n))}  |\lambda+\Box \rangle.
\end{eqnarray}

We can mimic the above calculation to end up with an analogous expression for $f_0$. Furthermore, inclusion of the $c_1(L)^m$ factor leads simply to adding a multiplicative factor given by the $m$'th power of $\epsilon_{\Box}=n_1\epsilon_1+n_2\epsilon_2+n_3\epsilon_3$ being the weight of $\mathbb{C}_{\epsilon_{\Box}}$ associated with the added box \cite{2011dd,Rapcak:2020ueh}. In total, we end up with formulas 
\begin{eqnarray}\nonumber
e_m|\lambda \rangle &=&\sum_{(\lambda+\Box,\lambda)\in F_{n+1,n}} \epsilon_{\Box}^m\frac{e_{U(1)^2}(T^*_{\lambda}(M(n))}{e_{U(1)^2}(T^*_{\lambda+\Box,\lambda}(M(n+1,n))}  |\lambda+\Box \rangle,\\
f_m|\lambda+\Box \rangle &=& \sum_{(\lambda+\Box,\lambda)\in F_{n+1,n}} \epsilon_{\Box}^m\frac{e_{U(1)^2}(T^*_{\lambda+\Box}(M(n+1))}{e_{U(1)^2}(T^*_{\lambda+\Box,\lambda}(M(n+1,n))}  |\lambda \rangle.
\label{ef}
\end{eqnarray}

To determine the action of generators $e_n,f_n$ in the fixed-point basis, we only need to find the ratio of Euler characters in the expressions above. Let us start with finding the decomposition of the the tangent space to $\mathcal{M}(m)$ at a fixed point $\lambda$ into representations of $U(1)^2$. At a given fixed point, we know from the above that the vector space $\mathbb{C}^n$ (and its dual) decomposes into weights specified by the Young diagram as
\begin{eqnarray}
V_{\lambda}= \oplus_{\Box \in \lambda}\mathbb{C}_{\epsilon_{\Box}},\qquad
V^*_{\lambda}= \oplus_{\Box \in \lambda}\mathbb{C}_{-\epsilon_{\Box}},
\end{eqnarray}
where we introduced the notation $V_{\lambda}$ for the vector space $\mathbb{C}^n$ together with the data specifying such a decomposition. Each $B_i$ is a map from $V_\lambda$ to $V_\lambda$ of weight $\epsilon_i$. It thus contributes to the character by $V_\lambda\otimes V_\lambda^*$ twisted by an extra factor of $\mathbb{C}^*_{\epsilon_i}$ that we are going to justify momentarily. Similarly, $I$ is a map from $\mathbb{C}$ to $V_\lambda$ and contributes by the factor of $V_{\lambda}$. Finally, each $J$ contributes by a factor of $V^*_{\lambda}$ tensored with $\mathbb{C}_\epsilon^*$ for some weight $\epsilon$ depending on the choice of the non-compact brane. See table \ref{weights} for details. In total, the arrows in the quiver diagram contribute by
\begin{eqnarray}
(\mathbb{C}_{\epsilon_1}^*+ \mathbb{C}_{\epsilon_2}^*+ \mathbb{C}_{\epsilon_3}^*)\otimes V_n\otimes V_{\lambda}^*+ V_{\lambda} + V_{\lambda} \otimes \mathbb{C}_{\epsilon}^*.
\end{eqnarray}
Dividing by the linearized gauge group leads to a subtraction of one factor of $V_{\lambda}\otimes V^*_{\lambda}$ producing
\begin{eqnarray}
T_{\lambda}=(\mathbb{C}_{\epsilon_1}^*+ \mathbb{C}_{\epsilon_2}^*+ \mathbb{C}_{\epsilon_3}^*-1)\otimes V_\lambda\otimes V_\lambda^*+ V_\lambda + V_\lambda \otimes \mathbb{C}_{\epsilon}.
\label{reff}
\end{eqnarray}

Let us justify the necessity to twist by $\mathbb{C}^*_{\epsilon_i}$ in the above expression (\ref{reff}). Let us look into the example of a single D2-brane oriented along $\mathbb{C}_{\epsilon_1}$ bound to two D0-branes, i.e. $n=2$. Forgetting about the equivariant $U(1)^2$-action, the tangent space of $M(2)$ at a given point is given by the cohomology of the complex
\begin{eqnarray}
\xymatrixcolsep{3pc}
\xymatrixrowsep{3pc}
\xymatrix{ \mbox{Hom}{[} \mathbb{C}^2,\mathbb{C}^2{]} \ar[r]  & \mathbb{C}^3 \otimes \mbox{Hom}{[}\mathbb{C}^2,\mathbb{C}^2{]} + \mbox{Hom}{[}\mathbb{C},\mathbb{C}^2{]} + \mathbb{C}^2\otimes  \mbox{Hom}{[}\mathbb{C}^2,\mathbb{C}{]}  }
\end{eqnarray}
with a differential 
\begin{eqnarray}
d= \begin{pmatrix}{[}B_1,\cdot{]}& {[}B_2,\cdot{]}& {[} B_3,\cdot{]}& \cdot I&J\cdot \end{pmatrix}^T
\end{eqnarray}
coming from the linearized gauge transformation. As we have seen in the previous section, $\mathbb{C}^2$ decomposes as a $U(1)^2$ representation as $\mathbb{C}_{0}\oplus \mathbb{C}_{\epsilon_1}$ and $B_1$ acts by
\begin{eqnarray}
B_{1}=\begin{pmatrix} 0&0 \\ 1 &0 \end{pmatrix}.
\end{eqnarray}
The commutation of $g\in \mbox{Hom}[ \mathbb{C}^2,\mathbb{C}^2]$ with $B_1$ thus maps
\begin{eqnarray}
\left [\begin{pmatrix} 0&0 \\ 1 &0 \end{pmatrix},\begin{pmatrix} \mathbb{C}_{0}&\mathbb{C}_{-\epsilon_1} \\ \mathbb{C}_{\epsilon_1} &\mathbb{C}_{0} \end{pmatrix}\right ] =\begin{pmatrix} \mathbb{C}_{-\epsilon_1}&0 \\ \mathbb{C}_{0} &\mathbb{C}_{-\epsilon_1} \end{pmatrix} \subset \mathbb{C}_{-\epsilon_1}\otimes \begin{pmatrix} \mathbb{C}_{0}&\mathbb{C}_{-\epsilon_1} \\ \mathbb{C}_{\epsilon_1} &\mathbb{C}_{0} \end{pmatrix},
\end{eqnarray}
justifying the shift by $\mathbb{C}^*_{\epsilon_1}$. One can analogously proceed to justify the shifts of $B_2,B_3,I$ and $J$.

Computation of the tangent space to the correspondence is just slightly more involved. Let us view the correspondence in terms of the embedding inside $M(n+1)\times M(n)$. The tangent space at a fixed point $(\lambda+\Box,\lambda)$ can be then written as\footnote{We have also introduced minus sign to denote factors $\mathbb{C}_{\epsilon}$ that get removed. Below, we are more generally label by minus sign all $\mathbb{C}_{\epsilon}$  factors that contribute by a term in denominator.}
\begin{eqnarray}
T_{\lambda+\Box,\lambda}=T_{\lambda}+T_{\lambda+\Box}-N_{\lambda+\Box,\lambda},
\label{reff2}
\end{eqnarray}
where $N_{\lambda+\Box,\lambda}$ is the normal bundle to the correspondence. But the normal bundle has the following decomposition as a $U(1)^2$ representation
\begin{eqnarray}
N_{\lambda+\Box,\lambda}=(\mathbb{C}_{\epsilon_1}^*+\mathbb{C}_{\epsilon_2}^*+\mathbb{C}_{\epsilon_3}^*-1)\otimes V_{\lambda}\otimes V_{\lambda+\Box}^*+ V_{\lambda} + V_{\lambda+\Box}^* \oplus\mathbb{C}_{\epsilon}^*.
\label{reff3}
\end{eqnarray}
This expression can be directly deduced from constrains (\ref{constraint}) defining the correspondence. More concretely, the $\mathbb{C}^*_{\epsilon_i}\otimes V_{\lambda}\otimes V^*_{\lambda+\Box}$ terms originate from the condition $\xi B^{(1)}_i=B^{(2)}_i \xi$, the term $V_{\lambda}$ from $\xi I^{(1)}=I^{(2)}$ and $V_{\lambda+\Box}^* \oplus\mathbb{C}_{\epsilon}^*$ from $J^{(1)}_i=J^{(2)}_{i}\xi$. Finally, $-V_{\lambda}\otimes V_{\lambda+\Box}^*$, that contributes positively to the tangent space due to the presence of two minuses, accounts for the added map $\xi$ itself.

Using (\ref{reff2}) in (\ref{ef}) and writing all the factors that contribute to the denominator with a minus sign, we can write
\begin{eqnarray}\nonumber
e_m|\lambda \rangle &=&\sum_{(\lambda+\Box,\lambda)\in F_{n+1,n}} \epsilon_{\Box}^m e_{U(1)^2}(N_{\lambda+\Box,\lambda}-T_{\lambda+\Box}) |\lambda+\Box \rangle,\\
f_m|\lambda+\Box \rangle &=& \sum_{(\lambda+\Box,\lambda)\in F_{n+1,n}} \epsilon_{\Box}^m e_{U(1)^2}(N_{\lambda+\Box,\lambda}-T_{\lambda})   |\lambda \rangle,
\end{eqnarray}
where $N_{\lambda+\Box,\lambda}$ and $T_{\lambda}$ are given by (\ref{reff}) and (\ref{reff3}). We have also implemented a notation where each $-\mathbb{C}_\epsilon$ factor with a minus sign contributes by $1/\epsilon$ into the character formula, i.e. a factor in denominator. Using
\begin{eqnarray}
\mathbb{C}_{\epsilon}\otimes \mathbb{C}_{\tilde{\epsilon}}=\mathbb{C}_{\epsilon+\tilde{\epsilon}},
\end{eqnarray}
we can simplify the expression for $N_{\lambda+\Box,\lambda}$ and $T_{\lambda}$ from (\ref{reff}) and (\ref{reff3}) to a sum of simple factors
\begin{eqnarray}
\sum_{\alpha}\mathbb{C}_{\epsilon_{\alpha}}-\sum_{\beta}\mathbb{C}_{\epsilon_{\beta}}
\end{eqnarray}
with the Euler character given by the ratio
\begin{eqnarray}
\left (\prod_{\alpha}\epsilon_{\alpha}\right )/\left (\prod_{\beta}\epsilon_{\beta}\right ).
\end{eqnarray}

Let us finish this section with a table showing an explicit form of the factor $\mathbb{C}_{\epsilon}$ for branes of different dimensions and orientations. This table can be easily deduced from the discussion in the section on flavor symmetries \ref{sec:flavor}:
\begin{center}
\begin{tabular}{| c | c c c | c c c |}\hline
 D6 & D4$_{x_1,x_2}$  &D4$_{x_1,x_3}$  &D4$_{x_2,x_3}$   &D2$_{x_1}$   &D2$_{x_2}$   &D2$_{x_3}$   \\  \hline
 $\emptyset$ & $\mathbb{C}_{\epsilon_1+\epsilon_2}$  & $\mathbb{C}_{\epsilon_1+\epsilon_3}$   &  $\mathbb{C}_{\epsilon_2+\epsilon_3}$  &   $\mathbb{C}_{\epsilon_1+\epsilon_2}+\mathbb{C}_{\epsilon_1+\epsilon_3}$ &   $\mathbb{C}_{\epsilon_1+\epsilon_2}+\mathbb{C}_{\epsilon_2+\epsilon_3}$   &   $\mathbb{C}_{\epsilon_1+\epsilon_3}+\mathbb{C}_{\epsilon_2+\epsilon_3}$ \\ \hline
\end{tabular}
\label{weights}
\end{center}

\subsection{D2-brane and the Weyl algebra}

Let us now explicitly evaluate the above expressions in the case of the D2-moduli space. Since we have a single fixed point associated with each dimension $n$, we are going to implement a simplified notation $|n\rangle$ for the fixed-point basis vector of $H^*_{U(1)^2}(M(n),\mbox{Crit}(W))$. According to the above formulas, for $f_m| n+1\rangle$, we get 
\begin{eqnarray}
(n\epsilon_1)^{m}\frac{e(V_n\otimes V_{n+1}^*\otimes (\mathbb{C}^*_{\epsilon_1}+\mathbb{C}^*_{\epsilon_2}+\mathbb{C}^*_{\epsilon_3}-1)+ V_{n}+(\mathbb{C}^*_{\epsilon_1+\epsilon_2}+\mathbb{C}^*_{\epsilon_1+\epsilon_3})\otimes V^*_{n+1})}{e(V_{n}\otimes V_{n}^*\otimes (\mathbb{C}_{\epsilon_1}^*+\mathbb{C}^*_{\epsilon_2}+\mathbb{C}_{\epsilon_3}^*-1)+ V_{n}+(\mathbb{C}^*_{\epsilon_1+\epsilon_2}+\mathbb{C}^*_{\epsilon_1+\epsilon_3})\otimes V^*_{n})}| n\rangle.
\label{aux1}
\end{eqnarray}
This can be simplified to
\begin{eqnarray}\label{exprd}
&(n\epsilon_1)^{m}(n\epsilon_1-\epsilon_3)(n\epsilon_1-\epsilon_2)\times \\ \nonumber
&\times \prod_{i=1}^{n}\frac{(n\epsilon_1-(i-1)\epsilon_1+\epsilon_1)(n\epsilon_1-(i-1)\epsilon_1+\epsilon_2)(n\epsilon_1-(i-1)\epsilon_1+\epsilon_3)}{n\epsilon_1-(i-1)\epsilon_1}| n\rangle.
\end{eqnarray}
Let us comment on the origin of various factors. The vector space $\mathbb{C}^n$ associated to the fixed point of $M(n)$ decomposes as $V_n=\mathbb{C}_0\oplus \mathbb{C}_{\epsilon_1}\oplus \dots \oplus \mathbb{C}_{(n-1)\epsilon_1}$. We then have an obvious cancellation 
\begin{eqnarray}
V_n\otimes V_{n+1}^*- V_{n}\otimes V^*_{n}&=& V_n\otimes (\mathbb{C}_0\oplus \dots \oplus \mathbb{C}_{-n\epsilon_1}-\mathbb{C}_0\oplus \dots \oplus \mathbb{C}_{-(n-1)\epsilon_1})\\ \nonumber
&=&V_n\otimes \mathbb{C}_{-n\epsilon_1}=\mathbb{C}_{-n\epsilon_1}\oplus \dots \oplus \mathbb{C}_{-\epsilon_1},
\end{eqnarray}
reproducing the terms $\prod_i (n\epsilon_1-(i-1)\epsilon_1)$ in the denominator (up to a sign) and accounting for the terms in the numerator and the denominator of (\ref{aux1}) with a negative sign. Similarly, twisting by $\mathbb{C}^*_{\epsilon_i}$ reproduces the terms in the numerator of the second line in (\ref{exprd}). The factors $V_n$ in the numerator and the denominator of (\ref{aux1}) cancel out. Using 
\begin{eqnarray} 
V^*_{n+1}-V_n^*=\mathbb{C}_{-n\epsilon_1}
\end{eqnarray}
and twisting by $\mathbb{C}^*_{\epsilon_1+\epsilon_2}$ and $\mathbb{C}^*_{\epsilon_1+\epsilon_3}$ respectively and using the relation $\epsilon_1+\epsilon_2+\epsilon_3=0$ reproduces the factors $(n\epsilon_1-\epsilon_3)(n\epsilon_1-\epsilon_2)$. Finally, $(n\epsilon_1)^m$ is simply the $m$'th power of the weight of the vectors space associated with of the added box.

For $e_m|n\rangle$, we analogously get
\begin{eqnarray}
\frac{(n\epsilon_1)^{m}e(V_n\otimes V^*_{n+1}\otimes (\mathbb{C}^*_{\epsilon_1}+\mathbb{C}^*_{\epsilon_2}+\mathbb{C}^*_{\epsilon_3}-1)+ V_{n}+(\mathbb{C}^*_{\epsilon_1+\epsilon_2}+\mathbb{C}^*_{\epsilon_1+\epsilon_3})\otimes V^*_{n+1})}{e(V_{n+1}\otimes V_{n+1}^*\otimes (\mathbb{C}_{\epsilon_1}^*+\mathbb{C}^*_{\epsilon_2}+\mathbb{C}^*_{\epsilon_3}-1)+ V_{n+1}+(\mathbb{C}_{\epsilon_1+\epsilon_2}^*+\mathbb{C}^*_{\epsilon_1+\epsilon_3})\otimes V^*_{n+1})}|n+1\rangle
\end{eqnarray}
that becomes
\begin{eqnarray}
(n\epsilon_1)^{m-1}\prod_{i=1}^{n+1}\frac{n\epsilon_1-(i-1)\epsilon_1}{(n\epsilon_1-(i-1)\epsilon_1-\epsilon_1)(n\epsilon_1-(i-1)\epsilon_1-\epsilon_2)(n\epsilon_1-(i-1)\epsilon_1-\epsilon_3)}|n+1\rangle.
\end{eqnarray}
Many terms in the above products cancel out and we can simplify them as
\begin{eqnarray}\nonumber
f_m|n+1 \rangle &=&(n+1)(n\epsilon_1)^m\prod_{i=1}^{n+1}((n+1-i)\epsilon_1-\epsilon_2)((n+1-i)\epsilon_1-\epsilon_3)|n \rangle, \\ \nonumber
e_m|n \rangle &=&-\frac{1}{\epsilon_1}(n\epsilon_1)^{m}\prod_{i=1}^{n+1}\frac{1}{((n+1-i)\epsilon_1-\epsilon_2)((n+1-i)\epsilon_1-\epsilon_3)}|n+1 \rangle.
\end{eqnarray}
Let us introduce
\begin{eqnarray}\nonumber
A_k=-\prod_{i=1}^{k}\frac{1}{((k+1-i)\epsilon_1-\epsilon_2)((k+1-i)\epsilon_1-\epsilon_3)}
\end{eqnarray}
and renormalize
\begin{eqnarray}\nonumber
 \prod_{k=1}^{m}A_k|m\rangle \rightarrow |m\rangle .
\end{eqnarray}
In terms of the renormalized basis, we have
\begin{eqnarray}\nonumber
f_n|m+1 \rangle &=&-(m+1)(\epsilon_1m)^n|m \rangle,\\ \nonumber
e_n|m \rangle &=&\frac{1}{\epsilon_1}(\epsilon_1m)^{n}|m+1 \rangle
\end{eqnarray}
and if we identify
\begin{eqnarray}\nonumber
|m\rangle =z^m,
\end{eqnarray}
the above action factors through 
\begin{eqnarray}\nonumber
f_n&\rightarrow& -(\epsilon_1z\partial)^n \partial,\\ \nonumber
e_n&\rightarrow& \frac{1}{\epsilon_1}z(\epsilon_1z\partial)^n
\end{eqnarray}
acting on $\mathbb{C}[z]$. In particular,
\begin{eqnarray}\nonumber
f_n |m+1\rangle &\rightarrow& -(\epsilon_1z\partial)^n \partial z^{m+1}=-(m+1)(\epsilon_1z\partial)^n  z^m= -(m+1)(\epsilon_1m)^nz^m\\ \nonumber
 &\rightarrow&  -(m+1)(\epsilon_1m)^n |m\rangle,\\
e_n |m\rangle &\rightarrow& \frac{1}{\epsilon_1}z(\epsilon_1z\partial)^n  z^{m}=\frac{1}{\epsilon_1}(\epsilon_1m)^n  z^{m+1}\rightarrow  \frac{1}{\epsilon_1}(\epsilon_1m)^n |m\rangle.
\end{eqnarray}
We have ended up with a geometric construction of the so-called vector representation of the 1-shifted affine Yangian \cite{2017ee,BR,RSYZ}.

\subsection{D4-brane and the $\widehat{\mathfrak{gl}}_1$ Kac-Moody algebra}

One can analogously perform the calculation for the D4-brane framing to obtain the so-called Fock representation of the affine Yangian \cite{nakajima,Nakajima:1994nid,2011dd,2017ee,2019}. Instead of going through the algebra, let us simply state the result. 

Let us introduce an associative algebra generated by $J_n$ for $n\in \mathbb{Z}$ and satisfying commutation relations\footnote{For a D4-brane of a different orientation, we would choose a normalization coming simply from the permutation of parameters $\epsilon_1,\epsilon_2,\epsilon_3$.}
\begin{eqnarray}
[J_m,J_n]=-\frac{1}{\epsilon_1\epsilon_2}m\delta_{m,-n}.
\end{eqnarray}
This algebra is known as the $\widehat{\mathfrak{gl}}_1$ Kac-Moody algebra, the $\mathfrak{gl}_1$ current algebra or the Heisenberg vertex operator algebra. It turns out that this algebra can be given a structure of a vertex operator algebra but this point is not going to be important for our discussion. Generally, configurations of D4-branes are always expected to lead to a vertex operator algebra leading to an interesting interppay between the theory of VOAs and geometry of divisors in Calabi-Yau threefolds \cite{Prochazka:2017qum,Rapcak:2018nsl}. 

The  $\widehat{\mathfrak{gl}}_1$ Kac-Moody algebra admits a class of lowest-weight modules generated by the action of negative modes $J_n$ on the lowest-weight state $|\mu\rangle $ satisfying
\begin{eqnarray}
J_0|\mu\rangle=\mu|\mu\rangle,\qquad J_m|\mu\rangle=0\qquad \mbox{for } m>0.
\end{eqnarray}
On top of the lowest-weight state $|\mu\rangle$, we can start building a pyramid of states by the action of negative modes
\begin{eqnarray} \nonumber
&|\mu\rangle& \\ \nonumber
&J_{-1}|\mu\rangle& \\
&J_{-1}^2|\mu\rangle,\quad J_{-2}|\mu\rangle &\\  \nonumber
&J_{-1}^3|\mu\rangle,\quad J_{-1}J_{-2}|\mu\rangle,\quad J_{-3}|\mu\rangle &
\end{eqnarray}
and so on. Each line corresponds to a subspace of a fixed weight given by the sum of indices associated with all the involved $J_n$ generators. Note also that these expressions are naturally in correspondence with 2d partitions:
\begin{eqnarray}\nonumber
&\varnothing& \\ \nonumber
&\yng(1)&\\
&\yng(1,1)\qquad \yng(2)&\\ \nonumber
&\yng(1,1,1)\qquad \yng(1,2)\qquad \yng(3)&
\end{eqnarray}

Let us introduce an alternative basis of the affine Yangian of $\mathfrak{gl}_1$ given by
\begin{eqnarray}
\tilde{J}_{-n}=\frac{1}{(m-1)!}\mbox{ad}_{e_1}^{m-1}e_0,\qquad \tilde{J}_{n}=-\frac{1}{(m-1)!}\mbox{ad}_{f_1}^{m-1}f_0,\qquad T_{2,0}={[}f_1,f_2{]}.
\end{eqnarray}
It turns out that the geometric action constructed above actually factors through
\begin{eqnarray}
Y_{0,0,1}&:&\tilde{J}_n\rightarrow J_n,\\ \nonumber
Y_{0,0,1}&:&T_{2,0}\rightarrow \frac{\epsilon_1^2\epsilon_2^2}{3}\sum_{k,m=-\infty}^{\infty}:J_{-m-k-2}J_{m}J_{k}:+\frac{\epsilon_1\epsilon_2\epsilon_3}{2}\sum_{m=1}^{\infty}m J_{-m-1}J_{m-1} 
\end{eqnarray}
acting on the above Fock module for $\mu=0$. For details see \cite{nakajima,Nakajima:1994nid,2011dd,2017ee,2019,Rapcak:2018nsl}. Permuting parameters $\epsilon_i$, we obtain two more representations $Y_{1,0,0}$ and $Y_{0,1,0}$ associated with the other two orientations of the D4-brane.

It is also possible to recover the general $\mu$ by introducing an equivariant parameter associated with the $GL(1)$ action on the vector space associated with the framing node. This refinement turns out to be essential for understanding representations associated with more complicated configurations of D4-branes as we are going to sketch in the next section.

\subsection{D6-brane and the MacMahon module}

Analogously, one can construct a representation of a -1-shifted affine Yangian on a vector space labeled by 3d partitions associated with  the D6-brane framing. It is straightforward to find explicit relations and recover the formulas\footnote{The action of such a -1 affine Yangian is presumably equivalent to the one introduced in \cite{FJMM,Prochazka:2015deb} if we restrict to the unshifted-Yangian subalgebra and renormalize our states.} form \cite{Rapcak:2020ueh}.

\section{Cherednik algebras and $\mathcal{W}$-algebras}

This section is an exploration of various modules one encounters when introducing stacks of non-compactly supported branes, non-compact branes that mutually intersect and nilpotent vacuum expectation value for Higgs fields.

\subsection{Affine Yangian of $\mathfrak{gl}_1$ and its shifts}

One can show that generators $e_m,f_m,\psi_m$ coming from any choice of the framing satisfy relations of the $\mathfrak{gl}_1$ affine Yangian or its shifted versions. The relations of the $\mathfrak{gl}_1$ affine Yangian are \cite{Maulik:2012wi,schiffmann2012cherednik,2017ee}
\begin{eqnarray}\nonumber
\psi_{i+j}&=&{[}e_i,f_j{]},\qquad {[}\psi_i,\psi_j{]}=0,\\ \nonumber
0&=&{[}e_{i+3},e_j{]}-3{[}e_{i+2},e_{j+1}{]}+3{[}e_{i+1},e_{j+2}{]}-{[}e_{i},e_{j+3}{]}\\ \nonumber
&&+\sigma_2 {[}e_{i+1},e_j{]}-\sigma_2{[}e_{i},e_{j+1}{]}-\sigma_3 \{ e_{i},e_j\},\\ \nonumber
0&=&{[}f_{i+3},f_j{]}-3{[}f_{i+2},f_{j+1}{]}+3{[}f_{i+1},f_{j+2}{]}-{[}f_{i},f_{j+3}{]}\\ \nonumber
&&+\sigma_2 {[}f_{i+1},f_j{]}-\sigma_2{[}f_{i},f_{j+1}{]}+\sigma_3 \{ f_{i},f_j\},\\ \nonumber
0&=&{[}\psi_{i+3},e_j{]}-3{[}\psi_{i+2},e_{j+1}{]}+3{[}\psi_{i+1},e_{j+2}{]}-{[}\psi_{i},e_{j+3}{]}\\ \nonumber
&&+\sigma_2 {[}\psi_{i+1},e_j{]}-\sigma_2{[}\psi_{i},e_{j+1}{]}-\sigma_3 \{ \psi_{i},e_j\},\\ \nonumber
0&=&{[}\psi_{i+3},f_j{]}-3{[}\psi_{i+2},f_{j+1}{]}+3{[}\psi_{i+1},f_{j+2}{]}-{[}\psi_{i},f_{j+3}{]}\\ 
&&+\sigma_2 {[}\psi_{i+1},f_j{]}-\sigma_2{[}\psi_{i},f_{j+1}{]}+\sigma_3 \{ \psi_{i},f_j\},
\label{yrelations2}
\end{eqnarray}
together with 
\begin{eqnarray}\nonumber
&{[}\psi_0,e_i{]}={[}\psi_0,f_i{]}={[}\psi_1,e_i{]}={[}\psi_1,f_i{]}=0,&\\
&{[}\psi_2,e_i{]}=2e_i,\qquad {[}\psi_2,f_i{]}=-2f_i&
\label{yrelations3}
\end{eqnarray}
and
\begin{eqnarray}\nonumber
\mbox{Sym}_{i,j,k}{[}e_i,{[}e_j,e_k{]}{]}&=&0,\\
\mbox{Sym}_{i,j,k}{[}f_i,{[}f_j,f_k{]}{]}&=&0,
\label{yrelations3}
\end{eqnarray}
where $\mbox{Sym}_{i,j,k}$ denotes symmetrization with respect to the indices. 

Its subalgebras generated by all $e_n,\psi_n$ but $f_m$ restricted to $m\geq k$ are called $k$-shifted affine Yangians. With a little bit of work, one can also introduce shifted Yangians with negative shift $k<0$ but let us not go into details. These can be found in \cite{Rapcak:2020ueh} for the $\mathfrak{gl}_1$ affine Yangian and in \cite{Galakhov:2021xum} for more general examples.

As proposed in \cite{Prochazka:2017qum,Rapcak:2020ueh,Gaiotto:2020dsq}, shifts of the algebra capture a lot of information about the non-compact branes whose bound state with compactly supported branes we are interested in. For example:
\begin{enumerate}
\item The MacMahon representation associated with a single D6-brane naturally forms a representation of the $-1$ shifted affine Yangian as discussed in \cite{Rapcak:2020ueh}. This representation is expected to be equivalent to the one proposed in \cite{Prochazka:2015deb} based on \cite{FJMM}.
\item The representations associated with various configurations of D4-branes form various representations of the $\mathfrak{gl}_1$ Yangian itself \cite{schiffmann2012cherednik,Maulik:2012wi,2011dd,Prochazka:2017qum,Rapcak:2018nsl,Chuang:2019qdz,BR}. The central generator $\psi_0$ controls the number of D4-branes of a given configuration  \cite{2017ee,Prochazka:2018tlo} by formula
\begin{eqnarray}
\psi_0 =-\frac{k_1}{\epsilon_2\epsilon_3}-\frac{k_2}{\epsilon_1\epsilon_3}-\frac{k_3}{\epsilon_1\epsilon_2}.
\end{eqnarray}
One needs to be slightly careful when generalizing these statements to more complicated geometries and associated more general affine Yangians \cite{Rapcak:2020ueh,Li:2020rij}. In more general Yangians, one has the choice to shift either the imaginary root or real roots \cite{Rapcak:2020ueh}. In the presence of only D4-brane-framings, the imaginary root is generally expected to remain unshifted but real roots carry an interesting information about the configuration of D4-branes. If we view a configuration of D4-branes as specifying a divisor inside our Calabi-Yau threefold (a sum of cycles on which our branes are supported together with multiplicities encoding the number of branes of a given support), the shifts are controlled by the intersection number of such a divisor with different $\mathbb{CP}^1$'s in our geometry\cite{Prochazka:2017qum,Rapcak:2020ueh}. These intersection numbers together with an analogue of the $\psi_0$ eigenvalue are expected to fully determine the divisor associated with our D4-brane configuration \cite{Prochazka:2017qum,Rapcak:2019wzw}. 
\item The representations associated with D2-branes are expected to lead to positive shifts of the affine Yangian \cite{Kodera:2016faj,Gaiotto:2020dsq}. This is in particular true for the vector representation associated with a single D2-brane that gives rise to a representation of the 1-shifted $\mathfrak{gl}_1$ affine Yangian as we are going to see bellow. Note that in order to get the one-shifted Yangian in terms of a subalgebra of the non-shifted Yangian, we need to relabel the generators from the previous section as $f_m\rightarrow f_{m+1}$. After such a relabeling, the $\psi_1$ generator is generally expected to control the number of D2-branes of a given orientation \cite{Gaiotto:2019wcc,Gaiotto:2020dsq} by formula
\begin{eqnarray}
\psi_1 =\frac{k_1}{\epsilon_1}+\frac{k_2}{\epsilon_2}+\frac{k_3}{\epsilon_3}
\end{eqnarray}
for $k_i$ D2-branes along $\mathbb{C}_{\epsilon_i}$.
\end{enumerate}

\subsection{Coproduct}

More complicated representations of the $\mathfrak{gl}_1$ affine Yangian can be obtain by utilizing the coproduct structure. In particular, the affine Yangian admits a coproduct 
\begin{eqnarray}
\Delta: \mathcal{Y}\rightarrow  \mathcal{Y}\otimes  \mathcal{Y}
\end{eqnarray}
given by formulas \cite{schiffmann2012cherednik,Maulik:2012wi,2011dd,Gaiotto:2020dsq}
\begin{eqnarray}\nonumber
\Delta&:&\tilde{J}_{n}\rightarrow \tilde{J}_{n}\otimes \mathds{1}+\mathds{1}\otimes \tilde{J}_{n},\\
\Delta&:& T_{2,0}\rightarrow T_{2,0}\otimes \mathds{1} +\mathds{1}\otimes T_{2,0}+\epsilon_1\epsilon_2\epsilon_3\sum_{m=1}^{\infty}m\tilde{J}_{-m-1}\otimes \tilde{J}_{m-1}.
\label{copr}
\end{eqnarray}
In this section, we are going to use this coproduct to construct representations associated with more complicated configurations of branes. To show that the resulting modules agree with those constructed geometrically from the cohomology of our quiver moduli spaces would require some extra work. We refer an interested reader to the original literature.

\subsection{Corner vertex operator algebras}

We are now going to discuss how various configurations of D4-branes recover elements of representation theory of $\mathcal{W}$-algebras arising from truncations of the affine Yangian.

Let us first compose the coproduct with the two elementary representations $Y_{0,0,1}$ acting on the Fock spaces $\mathcal{F}_{\mu_1}\otimes \mathcal{F}_{\mu_2}$, i.e. let us introduce an action of an affine-Yangian generator $t$ via
\begin{eqnarray}
(Y_{0,0,1}\otimes Y_{0,01})\circ \Delta (t).
\end{eqnarray}
The states of $\mathcal{F}_{\mu_1}\otimes \mathcal{F}_{\mu_2}$ are in correspondence with a pair of partitions that are in turn in correspondence with fixed points of the quiver moduli with rank-two framing if we introduce equivariant parameters $\mu_1,\mu_2$ associated with the Cartan of $GL(2)$ acting on the vector space associated with the framing node \cite{schiffmann2012cherednik,2017ee}.

One can also show that the above map produces only a subalgebra of the tensor product of $\widehat{\mathfrak{gl}}_1$ Kac-Moody algebras \cite{schiffmann2012cherednik} know as the Virasoro algebra tensored with a singe copy of the  $\widehat{\mathfrak{gl}}_1$ Kac-Moody algebra \cite{Fateev:1987zh}, i.e. an algebra generated by $L_m,J_n$ such that
\begin{eqnarray}\nonumber
{[}J_m,J_n{]}&=&-\frac{2}{\epsilon_1\epsilon_2}\delta_{m,-n},\\
{[}L_m,J_n{]}&=&-nJ_{m+n},\\ \nonumber
{[}L_m,L_n{]}&=&(m-n)L_{m+n}+\frac{1}{6}\left (7+3\left (\frac{\epsilon_1}{\epsilon_2}+\frac{\epsilon_2}{\epsilon_1}\right )\right )n(n^2-1)\delta_{m,-n}.
\end{eqnarray}
The lowest-weight state then satisfies
\begin{eqnarray}
J_{m}|\mu_1,\mu_2\rangle =L_{m}|\mu_1,\mu_2\rangle  =0\quad \mbox{for } m>0
\end{eqnarray}
and it is an eigenstate of $J_0,L_0$ with eigenvalues being a function of equivariant parameters $\mu_1,\mu_2,\epsilon_1,\epsilon_2$. The higher-weight states are generated by the action of negative modes $L_{-m},J_{-m}$. 

For special values of $\mu_1,\mu_2$, the above-constructed module is not irreducible. For example, specializing $\mu_1,\mu_2$ such that  $J_{0}|\mu_1,\mu_2\rangle =L_{0}|\mu_1,\mu_2\rangle =0$, we can define an irreducible module imposing further constraint (see e.g. \cite{Ginsparg1988AppliedCF})
\begin{eqnarray}
L_{-1}|\mu_1,\mu_2\rangle =0.
\end{eqnarray}
The states of a given weight are now obviously counted by
\begin{eqnarray}
\prod_{n=0}^{\infty}\frac{1}{(1-q^{1+n})(1-q^{2+n})}=1+q+3q^2+5q^3+10q^4+\dots
\label{gencount}
\end{eqnarray}
compared to the original
\begin{eqnarray}
\prod_{n=0}^{\infty}\frac{1}{(1-q^{1+n})^2}=1+2q+5q^2+10q^3+20q^4+\dots.
\end{eqnarray}
The generating function (\ref{gencount}) counts so-called nested partitions. These are pairs of partitions $(\lambda_1,\lambda_2)$ such that $\lambda_2$ can be placed on top of $\lambda_1$ to form a 3d partition. Alternatively, these can be identified with 3d partitions that are restricted to live inside a sandwich of height 2 along one of the directions. This module turns out to have a geometric realization coming from turning on nilpotent vacuum expectation value to the Higgs field on D4-branes as discussed in \cite{Chuang:2019qdz}.

One can proceed with a construction of more complicated algebras associated with a generic configuration of D4-branes by using the coproduct $k_1+k_2+k_3-1$ times and then composing with\footnote{Note that this construction is manifestly invariant under the triality that permutes the $\mathbb{C}_{\epsilon_i}$ inside our Calabi-Yau threefold $\mathbb{C}_{\epsilon_1}\times \mathbb{C}_{\epsilon_2}\times\mathbb{C}_{\epsilon_3}$. If we were to utilize various string-theory dualities, we could relate this geometric picture to a very differently-looking configuration of brane webs. When viewed from such a different  perspective, this triality, extending the well-known Feigin-Frenkel duality \cite{Feigin:1990pn}, has many highly non-trivial consequences \cite{Gaiotto:2017euk}. See also \cite{Gaberdiel:2012ku,Prochazka:2017qum,Creutzig:2020zaj}.}
\begin{eqnarray}
Y_{1,0,0}^{\oplus k_1}\otimes Y_{0,1,0}^{\oplus k_2}\otimes Y_{0,0,1}^{\oplus k_3}.
\end{eqnarray}
This leads to a class of corner vertex operator algebras \cite{Gaiotto:2017euk} acting on the tensor product of $k_1+k_2+k_3$ Fock modules \cite{2017ee,Prochazka:2018tlo,pit,Litvinov:2016mgi,Jafferis:2006ny}. With a little bit of work, one can show that the geometric action coming from the quiver with three D4-brane framings factors through this map \cite{Rapcak:2018nsl}. 

Similarly as above, a particular specialization of equivariant parameters associated to the framing node leads to a reducible module. The irreducible quotient can be then identified with a module with states counted by pit partitions\footnote{A pit partition is a 3d partition that can fit under the corner shifted by vector $(k_1,k_2,k_3)$ from the origin. Remember that numbers $(k_1,k_2,k_3)$  count branes of a given orientation.} \cite{pit,Gaiotto:2017euk,Prochazka:2017qum} and with a quiver-quantum-mechanics origin again coming from turning on Higgs vacuum expectation value to fields on D4-branes \cite{BR}.

\subsection{Cherednik algebras and Calogero-Moser integrable systems}

Let us finish with an exploration of much-less-understood algebras associated with more general configurations of D2-branes. 

First, let us look into the representation arising form a single D2-brane along the direction $\mathbb{C}_{\epsilon_1}$ and let us start with identifying the representation of generators in the $\tilde{J}_n,{[}f_0,f_1{]}$ basis. We have
\begin{eqnarray}
f_1=-\partial,\quad f_2=-\epsilon_1z\partial^2,\quad  e_0=\frac{1}{\epsilon_1}z,\quad e_1=z^2\partial,
\end{eqnarray}
where we have implemented the shift $f_n\rightarrow f_{n+1}$ to view the one-shifted affine Yangian as a subalgebra of the standard affine Yangian. Let us extend this representation by introducing 
\begin{eqnarray}
f_0=-\frac{1}{\epsilon_1}\frac{1}{z},\quad  f_1=-\partial,\quad f_2=-\epsilon_1z\partial^2,\quad  e_0=\frac{1}{\epsilon_1}z,\quad e_1=z^2\partial.
\end{eqnarray}
One can easily show that this is a representation of the non-shifted affine Yangian\footnote{This representation is usually referred to as the vector representation \cite{2017ee}.} acting on $\mathbb{C}[z,1/z]$. This extension is necessary in order to be able to use the coproduct of the affine Yangian to give a proposal for the general D2-brane algebras.

We can then identify the generators
\begin{eqnarray}
J_n\rightarrow \frac{1}{\epsilon_1}z^n,\qquad {[}f_0,f_1{]} \rightarrow \epsilon_1\partial^2
\end{eqnarray}
used in the definition of the coproduct (\ref{copr}). Using the coproduct and composing with $A_{1,0,0}\otimes A_{1,0,0}$ leads to
\begin{eqnarray}\nonumber
\tilde{J}_n&\rightarrow& \frac{1}{\epsilon_1}(z^n_1+z^n_2),\\
{[}f_0,f_1{]}&\rightarrow& \epsilon_1\partial^2_1+\epsilon_1\partial^2_2+\epsilon_1\epsilon_2\epsilon_3\sum_{m=1}^{\infty}m \frac{z_1^{-m-1}}{\epsilon_1}\frac{z_2^{m-1}}{\epsilon_1}\\ \nonumber
&\rightarrow& \epsilon_1\partial^2_1+\epsilon_1\partial^2_2+\frac{\epsilon_2\epsilon_3}{\epsilon_1}\frac{2}{(z_1-z_2)^2},
\end{eqnarray}
where we omitted the tensor product and $\mathds{1}$ in the formula for the coproduct and resummed the infinite sum of $(z_2/z_1)^m$. These expressions are known to form the Dunkel representation of the Cherednik algebra \cite{Kirillov1997LecturesOA,Etingof_symplecticreflection} associated with $\mathfrak{gl}(2)$. The resulting algebras conjecturaly act on moduli spaces associated with D2-branes \cite{Gaiotto:2020dsq,BR} but very little is known about the geometric construction of D2-brane modules. 

Similarly, one can use the coproduct $k_1+k_2+k_3-1$ times and compose the result with
\begin{eqnarray}
A_{1,0,0}^{\oplus k_1}\otimes A_{0,1,0}^{\oplus k_2}\otimes A_{0,0,1}^{\oplus k_3},
\end{eqnarray}
leading to
\begin{eqnarray}\nonumber
\tilde{J}_{0,n} &\rightarrow&  \epsilon_1^{-1} \sum_{i=1}^{k_1} z_i^n + \epsilon_2^{-1} \sum_{i=1}^{k_2} (z'_i)^n + \epsilon_3^{-1} \sum_{i=1}^{k_3} (z''_i)^n,  \\ \nonumber
T_{2,0} &\rightarrow&  \epsilon_1 \sum_{i=1}^{k_1} \partial_{z_i}^2+ \frac{\epsilon_2 \epsilon_3}{\epsilon_1} \sum_{i<j} \frac{2}{(z_i-z_j)^2}+ \epsilon_1 \sum_{i,j} \frac{2}{(z'_i-z''_j)^2} +\\
&&+ \epsilon_2 \sum_{i=1}^{k_2} \partial_{z'_i}^2+ \frac{\epsilon_1 \epsilon_3}{\epsilon_2} \sum_{i<j} \frac{2}{(z'_i-z'_j)^2}+ \epsilon_2 \sum_{i,j} \frac{2}{(z_i-z''_j)^2} +\\ \nonumber
&&+ \epsilon_3 \sum_{i=1}^{k_3} \partial_{z''_i}^2+ \frac{\epsilon_1 \epsilon_2}{\epsilon_3} \sum_{i<j} \frac{2}{(z''_i-z''_j)^2}+ \epsilon_3 \sum_{i,j} \frac{2}{(z_i-z'_j)^2}.
\label{calogero}
\end{eqnarray}
We arrive at a three-parametric generalization of the Cherednik algebra from \cite{Gaiotto:2020dsq} and extending its two-parametric version from \cite{Langmann,Estienne:2011qk,unknown,defq,Nekrasov:2017gzb}. The algebras \ref{calogero} form a triality-covariant\footnote{ Note that the triality symmetry from \cite{etingof2020new} discovered in the context of the double current algebra, an algebra isomorphic to the 1-shifted affine Yangian, has a natural geometric justification \cite{Gaiotto:2020dsq} in terms of permutation of coordinate lines in our Calabi-Yau threefold $\mathbb{C}^3$.} class of representations of 1-shifted (or non-shifted) affine Yangian of $\mathfrak{gl}_1$.

\section*{Acknowledgement}
I would like to express my deepest gratitude to Dylan Butson, Yan Soibelman, Alexander Tsymbaliuk, Yaping Yang and Gufang Zhao for patiently answering all my math questions. The research of M.R. was supported by NSF grant 1521446, NSF grant 1820912, the Berkeley Center for Theoretical Physics and the Simons Foundation.

\bibliography{refs}
\bibliographystyle{JHEP}

\end{document}